\documentclass[%
 reprint,
 amsmath,amssymb,
 aps,
]{revtex4-2}

\usepackage{graphicx}
\usepackage{subcaption}
\usepackage{dcolumn}
\usepackage{bm}

\usepackage{physics}
\usepackage{mathtools}
\usepackage{hyperref}

\usepackage{dcolumn}

\usepackage{xcolor}

\usepackage[autostyle]{csquotes}

\usepackage{makecell}

\begin{document}

\preprint{APS/123-QED}

\title{Extraction of different electronic contributions to transport properties of transition metals from first principles}

\author{I. S. Galtsov}
\author{D. V. Minakov}
\author{P. R. Levashov}
\affiliation{%
Joint Institute for High Temperatures, Izhorskaya 13 Building 2, Moscow 125412, Russia
}%
\affiliation{%
Moscow Institute of Physics and Technology, Institutskiy Pereulok 9, Dolgoprudny, Moscow Region 141701, Russia
}%

\date{\today}

\begin{abstract}
Accurate transport properties calculations in transition metals require treating inter-orbital hybridization, which is often poorly captured by simplified single-band models. We propose an \textit{ab initio} approach combining maximally localized Wannier functions and the Allen's method to correctly account for this mixing, overcoming the artifacts inherent to the Souza-Marzari-Vanderbilt scheme. Our method enables efficient orbital-resolved analysis of transport spectral functions. Applied to face-centered cubic Pd and body-centered cubic Mo, we reveal universal scattering mechanisms despite distinct electronic structures in these metals. Crucially, the parallel addition of distinct scattering channels perfectly reproduces full Allen's method calculations, rigorously validating this decomposition. We show that the absence of a $d$-peak at Fermi level in Mo enhances $s$-$s$ scattering, while the same contribution to resistivity and electron thermal conductivity of Pd is close to the experimental data for Ag. Ultimately, the proposed method provides a universal and efficient tool for the microscopic analysis of electron-phonon scattering mechanisms in systems with strong hybridization between localized and delocalized electron states.
\end{abstract}

\maketitle


\section{Introduction}\label{sec.intro}

Electron-phonon interaction is one of the central microscopic mechanisms governing the finite-temperature properties of metals. Although the subject has a long history \cite{ziman2001electrons, grimvall1981electron, Savrasov1996}, predictive calculations of electron-phonon coupling (EPC) and of the related transport coefficients have become broadly feasible only with the development of density-functional theory (DFT) theory \cite{GiustinoBook, BaroniDFPT, DeformPot2021, giri2019, GIRI2020100175}. Modern first-principles approaches now make it possible to compute electron-phonon matrix elements, relaxation rates and resistivities without relying on empirical scattering parameters \cite{giustino2017, EPW2023}. EPC is responsible for temperature dependence of thermoelectric properties \cite{Xu2020, XMa2023}, phonon linewidths \cite{Sklyadneva2006, sklyadneva2012} and the transfer of energy between electronic and ionic subsystems under nonequilibrium heating, including ultrafast excitation \cite{petrov2013, smirnov2020, smirnov2022, smirnov2025, smirnov2025Indium, akhmetov2023}.

Transition metals remain a particularly demanding class of systems for such calculations. Their transport properties are strongly affected by the coexistence of broad, highly dispersive $s$-like bands and narrow $d$-states close to the Fermi level \cite{Irkhin2021}. This electronic structure produces large and rapidly varying densities of states, band hybridization, and strongly anisotropic Fermi velocities. As a result, transition metals often display electrical resistivities and temperature dependences that cannot be understood using simple nearly free-electron models. A classic physical interpretation is provided by the two-band picture of $s$-$d$ scattering proposed by Mott, in which mobile conduction electrons are scattered into less mobile $d$-like states situated near the chemical potential \cite{mott1936electrical, mott1936resistance, mott1964electrons}. In this picture, the high density of $d$-states and their comparatively small velocities enhance the scattering probability and increase the resistivity. For phonon-limited transport, such processes are induced by phonon absorption or emission and are therefore naturally related to the electron-phonon matrix elements connecting electronic states of different orbital character.

The early theory of transition-metal resistivity established the basic physical language that is still useful today. Yamashita and Asano~\cite{yamashita1974} performed one of the first calculations of transition-metal resistivity based on realistic electronic structures and phonon spectra, showing that the actual Fermi-surface geometry, velocities, and wave-function character are essential for understanding Mo and Nb. Approximately at the same time Fradin~\cite{fradin1974} emphasized the role of sharp structure in the electronic density of states and showed that the anomalous temperature dependence of the resistivity of Pd can be reproduced only when the energy dependence of the density of states is retained rather than replaced by a constant Fermi-level value. These studies clarified that the usual simplifications appropriate for simple metals become questionable in systems such as Pd, where the $d$-band complex lies close to the Fermi level.

A major methodological step was made by Allen~\cite{allen_new}, who formulated the solution of the kinetic equation for electrons in metals in terms of special basis functions defined on the Fermi surface and in its vicinity. The lowest-order variational approximation (LOVA) exploited in this approach greatly reduces the computational cost while preserving the main physics for many metallic systems. Then Pinski, Allen, and co-workers~\cite{pinski1978, pinski1981} applied this approach to transition metals and demonstrated that realistic band structures and electron-phonon scattering probabilities can quantitatively describe the electrical and thermal resistivities of Nb and Pd. In the case of Pd, their results supported the $s$-$d$ scattering picture and showed that Fermi smearing plays a decisive role in the non-linear high-temperature behavior of the resistivity. This point is especially important because a low-temperature Fermi-surface approximation inevitably yields a high-temperature resistivity that is essentially linear in temperature and therefore misses the experimentally observed concavity in Pd~\cite{smirnov2022}.

In what follows, the work of Fradin~\cite{fradin1974} is of special interest as it represents an explicit attempt to evaluate the $s$--$d$ contribution to the electrical resistivity of palladium. The theoretical framework, based on the Ziman formalism~\cite{ziman2001electrons} with a Debye phonon spectrum, employed several adjustable parameters. Notably, the integration over electron energy was performed without invoking the commonly adopted---but strictly unjustified for palladium---assumption of a slowly varying electronic density of states. The intraband contribution $\rho_{ss}$, arising from all elementary scattering processes in which an electron initially occupying an $s$-like state is scattered into a vacant $s$-like state, was estimated by equating it to the experimental data on resistivity of silver.

The role of electronic states is also central in nonequilibrium metal dynamics. In ultrafast laser experiments, femtosecond excitation can raise the electronic temperature to several thousand kelvin while the lattice initially remains cold. The subsequent electron-to-lattice energy transfer is governed by EPC and is commonly modeled using the two-temperature model \cite{anisimov1974electron}. However, a single lattice temperature may be insufficient to describe early-time energy flow, motivating nonthermal or multi-temperature descriptions of the phonon system \cite{waldecker2016}. At the same time, multi-temperature models that distinguish different electronic subsystems, for example $s$- and $d$-electrons, require a microscopic separation of intra- and interband or intra- and interorbital contributions to the electron-phonon coupling strength \cite{medvedev2021contribution}.

It was shown by Petrov and Inogamov \cite{petrov2013} that strong femtosecond laser excitation can shift the chemical potential in transition metals above the $d$-band edge, thereby suppressing the Mott $s$--$d$ enhancement of the resistivity. This effect becomes important for electron temperatures of several eV, but it is not considered in this work as we focus on the equilibrium heating regime in the solid state.

Despite this progress, a first-principles interpretation of the Mott $s$-$d$ scattering mechanism remains nontrivial. In real transition metals the relevant bands are not purely $s$- or $d$-like. They are hybridized throughout substantial parts of the Brillouin zone, and the apparent separation into an $s$- band and a $d$-band complex depends on the chosen band representation, energy window, and disentanglement procedure. The method of Souza, Marzari, and Vanderbilt (SMV) \cite{souza2001} provides a rigorous and widely used framework for constructing maximally localized Wannier functions (MLWFs) for entangled band manifolds \cite{MLWF2012}. However, when the goal is not only to interpolate bands but also to assign scattering events to physically meaningful $s$- and $d$-related transport channels, the direct extraction of separate band manifolds may become ambiguous near avoided crossings and in regions of strong hybridization. This ambiguity is particularly serious for metals like Pd, where the $d$-states lie close to the Fermi level and are strongly mixed with more dispersive states.

The present work addresses this problem by developing a Wannier-projected analysis of electron-phonon coupling and phonon-limited electrical resistivity in transition metals. The central idea is to preserve the hybridized nature of the \textit{ab initio} electronic eigenstates while resolving their orbital content in a Wannier basis. Instead of forcing the band structure into separate pure $s$- and $d$-manifolds, we reconstruct the electronic energies and projection weights from a Wannier-interpolated representation and use these weights to decompose the electron-phonon spectral functions and transport properties into contributions associated with different orbital characters. It provides a natural first-principles interpretation of the Mott $s$-$d$ mechanism: scattering channels are identified not by an artificial band label but by the Wannier-projected orbital composition of the initial and final electronic states.


The purpose of this paper is therefore twofold. First, we provide a first-principles, Wannier-based formulation of $s$- and $d$-resolved electron-phonon transport in Pd and Mo that explicitly includes Fermi smearing and avoids empirical scattering parameters. Second, we demonstrate that the Mott $s$-$d$ scattering mechanism can be reinterpreted in terms of hybridized \textit{ab initio} electronic states, making it compatible with modern Wannier interpolation and with temperature-dependent transport and energy-relaxation calculations. This approach bridges the historical two-band interpretation of transition-metal resistivity with contemporary first-principles electron-phonon theory and provides a practical route for channel-resolved modeling of transport and ultrafast energy transfer in transition metals.

The paper is organized as follows. In sec.~\ref{sec.model} we first provide general formalism with respect to Allen's method for solving kinetic equation, then move to questions related to Wannier interpolation technique adopted, and exploit it to expand Allen's approach for accurate accounting for orbital hybridization. This section is finalized by focusing on important computational details specific to the methods proposed. Our results are discussed in sec.~\ref{sec:results} separately for case of Pd and Mo. The consclusions are given in ~\ref{sec.conclusion}.

\section{Formalism}\label{sec.model}

\subsection{Allen's method}\label{subsec.Allen}

This section presents the basic expressions of the method proposed by Allen \cite{allen_new} for calculating transport properties of metals. More details and examples can be found in other recent works \cite{smirnov2022, GALTSOV2025}. The main problem is to solve semiclassical kinetic linearized equation, which is written here for a small external electric field only:

\begin{equation}\label{eq:BTE}
\left(e\textbf{E} + \frac{\varepsilon_{n\textbf{k}}}{T} \nabla T \right) \textbf{\textit{v}}_{n\textbf{k}} \frac{\partial f_{n\textbf{k}}}{\partial \varepsilon_{n\textbf{k}}} = \sum_{m\textbf{k}^\prime} Q_{n\textbf{k}, m\textbf{k}^\prime} \phi_{m\textbf{k}^\prime}.
\end{equation}
The right-hand side of this equation contains the small additive $\phi_{n\textbf{k}}$ to the full nonequilibrium distribution function of electrons:

\begin{equation}\label{eq:DistFunc}
\varPhi_{n\textbf{k}} = f_{n\textbf{k}} + \left( - \frac{\partial f_{n\textbf{k}}}{\partial \varepsilon_{n\textbf{k}}} \right) \phi_{n\textbf{k}},
\end{equation}

\begin{equation*}\label{eq:FD}
f_{n\textbf{k}} = \frac{1}{e^\frac{\varepsilon_{n\textbf{k}} - \mu}{T} + 1},
\end{equation*}
where $f_{n\textbf{k}}$ is Fermi-Dirac function and $\mu$ is the chemical potential taken as the electronic energy zero in this work: $\mu = 0$. We also set Boltzmann constant $k_B = 1$ and measure temperatures in energy units. The kinetic equation written in the form \eqref{eq:BTE} depends on scattering operator matrix elements $Q_{n\textbf{k}, m\textbf{k}^\prime}$ to be precalculated. Generally, this matrix encapsulates different scattering mechanisms, the most important being finite-temperature lattice vibrations acting on conduction electrons, which is described in the EPC formalism. The small deviations $\phi_{n\textbf{k}}$ allow to explicitly evaluate electron and heat density currents (the factor of two is from summation over two spin components  -- we do not consider spin-polarized case in this work):

\begin{equation}\label{eq:el_current}
\textbf{j}_e =  -2 e \cdot \sum_{n\textbf{k}} \textbf{\textit{v}}_{n\textbf{k}} \phi_{n\textbf{k}} \left( -\frac{\partial f_{n\textbf{k}}}{\partial \varepsilon_{n\textbf{k}}} \right),
\end{equation}

\begin{equation}\label{eq:heat_current}
\textbf{j}_Q = 2 \cdot \sum_{n\textbf{k}} \varepsilon_{n\textbf{k}} \textbf{\textit{v}}_{n\textbf{k}} \phi_{n\textbf{k}} \left( -\frac{\partial f_{n\textbf{k}}}{\partial \varepsilon_{n\textbf{k}}} \right).
\end{equation}

The kinetic approach is based on the probability of an elementary electron scattering process:

\begin{multline}\label{eq:Pkk_for_elph}
P_{n\textbf{k},m\textbf{k}^\prime} = \dfrac{2\pi}{\hbar}  \sum_{\textbf{q} \nu} |g_{mn}^\nu(\textbf{k},\textbf{q})|^2 f(\varepsilon_{n\textbf{k}}) (1 - f(\varepsilon_{m\textbf{k}^\prime})) \\
\times 
[(n(\omega_{\textbf{q} \nu}, T) + 1)\delta(\varepsilon_{n\textbf{k}} - \varepsilon_{n\textbf{k}^\prime} - \omega_{\textbf{q} \nu} ) + \\
n(\omega_{\textbf{q} \nu}, T)\delta(\varepsilon_{n\textbf{k}} - \varepsilon_{n\textbf{k}^\prime} + \omega_{\textbf{q} \nu} )].
\end{multline}
This expression reflects Fermi's golden rule for a single electron with initial momentum $\textbf{k}$ and energy $\varepsilon_{n\textbf{k}}$ scattered due to a phonon with momentum $\textbf{q}$ and energy $\omega_{\textbf{q}\nu}$. In the final state the electron has momentum $\textbf{k}^\prime = \textbf{k} + \textbf{q}$ and energy $\varepsilon_{m\textbf{k}^\prime} = \varepsilon_{n\textbf{k}} \pm \omega_{\textbf{q}\nu}$ in accordance with the conservations laws given in delta functions. The statistics on phonons is given by Bose-Einstein distribution $n(\omega_{\textbf{q}\nu}, T) = (\exp(\omega_{\textbf{q}\nu}/T) - 1)^{-1}$.

The corresponding matrix element of the scattering operator in \eqref{eq:BTE} can be expressed in terms of the elementary probability defined above:

\begin{equation}\label{eq:scattering_op}
Q_{n\textbf{k},m\textbf{k}^\prime} = \frac{1}{T} \left( \delta_{\textbf{k}\textbf{k}^\prime} \delta_{nm} \sum_{l\textbf{k}^{\prime\prime}} P_{n\textbf{k},l\textbf{k}^{\prime\prime}} - P_{n\textbf{k},m\textbf{k}^{\prime}} \right).
\end{equation}

What was made by Allen in his seminal work is the change of basis in which the kinetic equation \eqref{eq:BTE} to be solved. For this purpose he presented the following basis functions:

\begin{equation}\label{eq:khi_Allen_basis}
\chi_{\alpha \varsigma}(n\textbf{k}) = \dfrac{F_\alpha(n\textbf{k}) \eta_\varsigma(\varepsilon_{n\textbf{k}})}{N(\varepsilon_{n\textbf{k}}) v(\varepsilon_{n\textbf{k}})},
\end{equation}
\begin{equation}\label{eq:gamma_Allen_basis}
\gamma_{\alpha \varsigma}(n\textbf{k}) = F_\alpha(n\textbf{k}) \eta_\varsigma(\varepsilon_{n\textbf{k}})  v(\varepsilon_{n\textbf{k}}) \left( - \dfrac{\partial f_{n\textbf{k}}}{\partial \varepsilon_{n\textbf{k}}} \right).
\end{equation}

In these expressions Fermi surface harmonics \cite{allen1976fermi} of the first order $F_\alpha(n\textbf{k}) = v_\alpha(n\textbf{k}) / v(\varepsilon_{n\textbf{k}})$ are exploited ($\alpha = x,y,z$ is the cartesian index). They are responsible for solving the angular part of the problem in the reciprocal space.  Here the normalization of electronic velocities is introduced:

\begin{equation}
v(\varepsilon) = \sqrt{\langle v_\alpha^2(\varepsilon) \rangle}, \;\;\; \langle v_\alpha^2(\varepsilon) \rangle = \frac{\sum_{n\textbf{k}} v_\alpha^2(n\textbf{k}) \delta(\varepsilon_{n\textbf{k}} - \varepsilon)}{\sum_{n\textbf{k}} \delta(\varepsilon_{n\textbf{k}} - \varepsilon)}.
\end{equation}
It is more convenient to explicitly denote the numerator and the denominator of the last expression. The denominator is simply electron density of states per spin (DOS):

\begin{equation}\label{eq:DOS}
N(\varepsilon) = \sum_{n\textbf{k}} \delta(\varepsilon_{n\textbf{k}} - \varepsilon),
\end{equation}
and the numerator is sometimes called the corresponding transport density of states (TrDOS) per spin:

\begin{equation}\label{eq:TrDOS}
N(\varepsilon)v_\alpha^2(\varepsilon) = \sum_{n\textbf{k}} v_\alpha^2(n\textbf{k}) \delta(\varepsilon_{n\textbf{k}} - \varepsilon),
\end{equation}
Only $xx$-component is considered further to simplify the notations, as we only work with cubic metals in this work and so the $\alpha$ index is omitted in velocities and TrDOS.

The second important part of Allen's formalism is connected to special energy-dependent polynomials which serve for solution of radial part of the problem. These polynomials are orthonormalized following

\begin{equation}\label{eq:allen_poly_ortho}
\int_{-\infty}^\infty \left( - \frac{\partial f}{\partial \varepsilon} \right) \eta_{\varsigma}(\varepsilon) \eta_{\varsigma^\prime}(\varepsilon) d\varepsilon.
\end{equation}
Working in LOVA only two first polynomials are used, namely $\eta_0 = 1$ and $\eta_1(\varepsilon) = \sqrt{3}\varepsilon / (\pi T)$.

Omitting extra derivations here, which can be found in papers \cite{allen_new,smirnov2022}, we write the general expression for scattering matrix in Allen's basis representation:

\begin{widetext}
  \begin{multline}\label{eq.QFull}
Q_{x \varsigma, x \varsigma^\prime} =
\dfrac{2 \pi V}{\hbar T}
\int_{0}^\infty d\Omega 
\int_{-\infty}^\infty d\varepsilon
\int_{-\infty}^\infty d\varepsilon^\prime
\sum_{s, s^\prime = \pm 1}
\alpha_\mathrm{tr}^2F(s,s^\prime, \varepsilon, \varepsilon^\prime, \Omega)
N(\varepsilon)
J(s,s^\prime, \varsigma, \varsigma^\prime \varepsilon, \varepsilon^\prime)
f(\varepsilon, T)
(1 - f(\varepsilon^\prime, T)) \times \\
[(n(\Omega, T) + 1)\delta(\varepsilon - \varepsilon^\prime - \Omega) + 
n(\Omega, T)\delta(\varepsilon - \varepsilon^\prime +\Omega)].
\end{multline}
Here $V$ is a unit cell volume. It is the central expression for the following. It consists of joint energy polynimials

\begin{equation}\label{eq:joint_poly}
J(s,s^\prime, \varsigma, \varsigma^\prime \varepsilon, \varepsilon^\prime) = \frac{1}{4} \left( \frac{\eta_{\varsigma}(\varepsilon)}{N(\varepsilon)v(\varepsilon)} + s \frac{\eta_{\varsigma}(\varepsilon^\prime)}{N(\varepsilon^\prime)v(\varepsilon^\prime)}\right) \left( \frac{\eta_{\varsigma^\prime}(\varepsilon)}{N(\varepsilon)v(\varepsilon)} + s^\prime \frac{\eta_{\varsigma^\prime}(\varepsilon^\prime)}{N(\varepsilon^\prime)v(\varepsilon^\prime)}\right)
\end{equation}
and the transport spectral function (TSF)

\begin{equation}
        \alpha_\mathrm{tr}^2(s,s^\prime, \varepsilon, \varepsilon^\prime, \Omega) = 
\frac{1}{2 N(\varepsilon)} \sum_{\textbf{q} \nu}
\delta(\Omega - \omega_{\textbf{q} \nu})
\sum_{mn,\textbf{k}}
|g_{mn}^\nu(\textbf{k},\textbf{q})|^2
\left[
F_{n\textbf{k}} - sF_{m\textbf{k}+\textbf{q}}
\right]
\left[
F_{n\textbf{k}} - s^\prime F_{m\textbf{k}+\textbf{q}}
\right] \delta(\varepsilon_{n\textbf{k}} - \varepsilon)
\delta(\varepsilon_{m\textbf{k}+\textbf{q}} - \varepsilon^\prime),
    \end{equation}
which encapsulates all the elementary scattering processes described above. Only the TSF with $s = s^\prime = 1$  is needed for electrical resistivity evaluation. Another important case is for $s = s^\prime = -1$, which contributes to thermal conductivity evaluation.

%


Following Smirnov \cite{smirnov2022} again, we further neglect the difference between the initial electronic energy $\varepsilon$ and the final one $\varepsilon^\prime$. Then the explicit expressions for TSFs become more feasible:

\begin{equation}\label{eq:A2F_pp}
\alpha^2_\mathrm{tr} F(1, 1, \varepsilon, \Omega) = \frac{1}{2N(\varepsilon)v^2(\varepsilon)} \sum_{\textbf{q}\nu} \delta(\Omega - \omega_{\textbf{q}\nu}) \sum_{mn,\textbf{k}} |g^\nu_{mn}(\textbf{k},\textbf{q})|^2
    (v_{n\textbf{k}} - v_{m\textbf{k} + \textbf{q}})^2 \delta(\varepsilon_{n\textbf{k}} - \varepsilon) \delta(\varepsilon_{m\textbf{k}+\textbf{q}} - \varepsilon),
\end{equation}

\begin{equation}\label{eq:A2F_mm}
\alpha^2_\mathrm{tr} F(-1, -1, \varepsilon, \Omega) = \frac{1}{2N(\varepsilon)v^2(\varepsilon)} \sum_{\textbf{q}\nu} \delta(\Omega - \omega_{\textbf{q}\nu}) \sum_{mn,\textbf{k}} |g^\nu_{mn}(\textbf{k},\textbf{q})|^2
    (v_{n\textbf{k}} + v_{m\textbf{k} + \textbf{q}})^2 \delta(\varepsilon_{n\textbf{k}} - \varepsilon) \delta(\varepsilon_{m\textbf{k}+\textbf{q}} - \varepsilon),
\end{equation}

The TSF in \eqref{eq:A2F_pp} allows one to find temperature dependence of electrical resistivity via the $\varsigma = \varsigma^\prime = 0$ scattering operator matrix element:

\begin{equation}\label{eq:rho_from_Q}
\rho(T) = \frac{1}{2e^2} Q_{x0,x0}(T).
\end{equation}
We explicitly write the expression for $Q_{x0,x0}$:


\begin{multline}\label{eq:ord0ExactFormulas}
Q_{x0,x0} =
\dfrac{2 \pi V}{\hbar T}
\int d\Omega
\int  d\varepsilon \dfrac{I_0(\varepsilon,\Omega)}{N(\varepsilon)v^2(\varepsilon)}, \\
I_0(\varepsilon,\Omega) = 
\alpha^2_{\mathrm{tr}} F(1,1,\varepsilon,\Omega)
\int d\varepsilon^\prime f(\varepsilon, T) (1 - f(\varepsilon^\prime, T)) \\ \times
[(n(\Omega, T) + 1)\delta(\varepsilon - \varepsilon^\prime - \Omega) + 
n(\Omega, T)\delta(\varepsilon - \varepsilon^\prime + \Omega)],
\end{multline}
The corresponding expressions for electron thermal conductivity read:

\begin{equation}\label{eq:kappa_from_Q}
\kappa_e = \frac{2\pi^2 T}{3} \left[ (Q)_{x1,x1}^{-1} - |(Q^{-1})_{x0,x1}|^2 Q_{x0,x0} \right],
\end{equation}

\begin{multline}\label{eq:ord1ExactFormulas}
Q_{x0,x1} =
\dfrac{\sqrt{3} V}{\hbar T^2}
\int d\Omega
\int  d\varepsilon \dfrac{I_{01}(\varepsilon,\Omega)}{N(\varepsilon)v^2(\varepsilon)}, \\
I_{01}(\varepsilon,\Omega) = 
\alpha^2_{\mathrm{tr}} F(1,1,\varepsilon,\Omega)
\int d\varepsilon^\prime (\varepsilon + \varepsilon^\prime) f(\varepsilon, T) (1 - f(\varepsilon^\prime, T)) \\ \times
[(n(\Omega, T) + 1)\delta(\varepsilon - \varepsilon^\prime - \Omega) + 
n(\Omega, T)\delta(\varepsilon - \varepsilon^\prime + \Omega)],
\end{multline}
\begin{multline}\label{eq:ord2ExactFormulas}
Q_{x1,x1} =
\dfrac{6 V}{\pi \hbar T^3}
\int d\Omega
\int  d\varepsilon \dfrac{I_1(\varepsilon,\Omega)}{N(\varepsilon)v^2(\varepsilon)}, \\
I_1(x,x,\varepsilon,\Omega) = 
\dfrac{1}{4} \sum_{s,s^\prime = \pm 1}\alpha^2_{\mathrm{tr}} F(s,s^\prime,\varepsilon,\Omega)
\int d\varepsilon^\prime  (\varepsilon + s\varepsilon^\prime)(\varepsilon + s^\prime \varepsilon^\prime) f(\varepsilon, T) (1 - f(\varepsilon^\prime, T)) \\ \times
[(n(\Omega, T) + 1)\delta(\varepsilon - \varepsilon^\prime - \Omega) + 
n(\Omega, T)\delta(\varepsilon - \varepsilon^\prime + \Omega)].
\end{multline}
Following Smirnov, we neglect the contributions with $s \neq s^\prime$ in \eqref{eq:ord2ExactFormulas} due to their minor contribution to the last expression.

\end{widetext}

\subsection{Wannier interpolation}\label{subsec.WannProj}

\subsubsection{Wannier-projected band structure}

In this section, we present the standard formalism used in the field of maximally localized Wannier functions. More details can be found in well-known publications and reviews \cite{MarzariVanderbilt1997, souza2001, Wang2006, Yates2007, Wang2007, Wannier90_2008, MLWF2012, SymWFs2013, agapito2018, vanderbilt2018berry, Wannier2024}. The main advantage of using MLWFs lies in the Wannier interpolation procedure, which enables highly accurate reconstruction of the electronic band structure of a solid-state material on very dense $\textbf{k}$-point grids. Moreover, Giustino and collaborators~\cite{giustino2007, giustino2017, EPW2016, EPW2023} have significantly extended this technique to reconstruct not only electronic energies but also electron–phonon coupling. The topic has already proven itself successfully and is still being developed quite actively~\cite{poliukhin2026carrier, WannierGPU2026}. Below, we briefly outline the key steps of Wannier interpolation, primarily to introduce the essential notation and to highlight important features specific to the case of transition metals.

In the most simple case, the subset of electronic states of interest forms an isolated group, meaning it is clearly separated in energy from all bands above and below. This situation is typical for insulators, where valence bands are separated from conduction bands by a gap. In such cases, the Marzari–Vanderbilt (MV) procedure~\cite{MarzariVanderbilt1997}, historically proposed first, works effectively.

When bands of interest are \enquote{entangled} with others, which is the case in transition metals in the vicinity of the chemical potential, one has to construct a small number of maximally localized Wannier functions that accurately describe them. The procedure was introduced by Souza, Marzari, and Vanderbilt (SMV)~\cite{souza2001}. In this approach a special energy window called disentanglement (or \enquote{outer}) window is exploited: at each point $\textbf{k}_c$ of a coarce initial grid in the first BZ the number of states $N_\mathrm{dis}(\textbf{k}_c)$ is determined by the lower $E_\mathrm{dis}^\mathrm{min}$ and upper $E_\mathrm{dis}^\mathrm{max}$ energy values which are adjustable parameters. The number  $N_\mathrm{sub} \le N_\mathrm{dis}(\textbf{k}_c)$ of states to be obtained from this \enquote{outer} manifold (e.g., $5$ states of the $d$-complex) is another input parameter same for all $\textbf{k}_c$. The goal is to contruct the new $N_\mathrm{sub}$-dimensional subspace of states (called optimal subspace) via an unitary transformation which relates them to the original set of $N_\mathrm{dis}(\textbf{k}_c)$ states:

\begin{equation}\label{eq:Bloch_in_W_gauge}
\ket{u^{\mathrm{sub}}_{n\textbf{k}_c}} = \sum_{m \in \mathrm{dis}(\textbf{k}_c)} \ket{u_{n\textbf{k}_c}} \mathcal{U}_{mn}^\mathrm{dis}(\textbf{k}_c).
\end{equation}
This transformation fulfills a critical role in \enquote{ironing out}~\cite{vanderbilt2018berry} the periodic parts $\ket{u_{n\textbf{k}_c}}$ of the original Bloch states, thereby removing spurious singularities that may arise at high-symmetry points within the Brillouin zone. The matrix $\mathcal{U}_{mn}^\mathrm{dis}(\textbf{k}_c)$ of shape $N_\mathrm{dis}(\textbf{k}_c) \times N_\mathrm{sub}$ is evaluated by the minimising the gauge-invariant part $\Omega_I$ of the spread functional $\Omega$ as given by SMV algorithm and then the gauge-dependent part $\tilde{\Omega}$ by MV technique. The set of $N_\mathrm{sub}$ MLWFs is obtained from these states via a Fourier transform

\begin{equation}\label{eq:WFs_from_Bloch}
\ket{n\textbf{R}_e} = \frac{1}{N_c^e} \sum_{\textbf{k}_c} e^{-i \textbf{k}_c \textbf{R}_e} \ket{u_{n\textbf{k}_c}^\mathrm{sub}}
\end{equation}
and actually describes the same optimal subspace as Bloch representation does. Here $N_c^e$ is the total number of \textbf{k}-points in the initial (DFT) coarse grid used for electronic structure calculations, $\ket{n\textbf{R}_e}$ is the $n$-th Wannier function located in a unit-cell shifted at $\textbf{R}_e$ from the home cell ($\textbf{R}_e = \textbf{0}$). The subscript \enquote{e} is chosen in the spirit of Giustino's original work~\cite{giustino2007} to distinguish this point in the real-space Wingner-Seitz supercell from the one related to phonons described below.

Since band mixing in the optimal subspace can distort original energy dispersions, an inner (or \enquote{frozen}) window is usually introduced to strictly preserve system properties within a critical energy range (e.g., near the Fermi level); states falling inside this frozen window (defined by two parameters $E_\mathrm{froz}^\mathrm{min}$ and $E_\mathrm{froz}^\mathrm{max}$) are included unchanged in the final subspace. This is particularly critical in cases of strong hybridization, such as when a broad $s$-type band intersects a narrow $d$-complex in transition metals. However, it is important to note that for a fundamentally different task—namely, the direct extraction of the $d$-complex (a five-dimensional optimal subspace)—the frozen window cannot be employed, as the resulting set of $d$-bands would differ from those predicted by DFT calculations.

Next, the Hamiltonian in Wannier representation is evaluated:

\begin{multline}\label{eq.HamWan2Bloch}
H_{nm}(\textbf{R}_e) = \bra{{n\textbf{0}}} \hat{H} \ket{{m\textbf{R}_e}} = \\ \sum_{\textbf{k}_c} e^{-i\textbf{k}_c \cdot \textbf{R}_e} [\mathcal{U}_\mathrm{dis}^\dagger(\textbf{k}_c) H(\textbf{k}_c)\mathcal{U}_\mathrm{dis}(\textbf{k}_c)]_{nm},
\end{multline}
where the Hamiltonian in the Bloch representation is diagonal: $H(\textbf{k}_c) = \delta_{mn}\varepsilon_{n\textbf{k}_c}$. One of the main restrictions imposed on the Wannier orbitals is their strong spatial localization in the direct space. Then the electronic energies $\varepsilon_{n\textbf{k}_f}$, obtained on fine grid $\textbf{k}_f$ are smooth functions of a \textbf{k}-vector. To evaluate them, the Hamiltonian in the Bloch representation on a fine grid is found via the inverse transformation:

\begin{equation}\label{eq.HamBloch2Wan}
\tilde{H}_{nm}(\textbf{k}_f) = \frac{1}{N_c^e} \sum_{\textbf{R}_e} e^{i \textbf{k}_f \textbf{R}_e} H_{nm}(\textbf{R}_e).
\end{equation}
The diagonalization of this matrix provides the final interpolated band energies at the specified $\textbf{k}_f$ point:

\begin{equation}\label{eq.diagHam}
\tilde{H}(\textbf{k}_f) = U(\textbf{k}_f) H(\textbf{k}_f) U^\dagger(\textbf{k}_f).
\end{equation}
Notice the different designation for the rotation matrix $U(\textbf{k}_f)$ as it has a significantly different meaning than the matrix $\mathcal{U}^\mathrm{dis}(\textbf{k}_c)$ introduced in \eqref{eq:Bloch_in_W_gauge}. The matrix $U(\textbf{k}_f)$ is a square matrix of shape $N_\mathrm{dis} \times N_\mathrm{dis}$, as it provides the interpolated electron energies in the optimal subspace: $H(\textbf{k}_f) = \delta_{mn}\varepsilon_{n\textbf{k}_f}$. Actually, the $\varepsilon_{n\textbf{k}_f}$ coincides with the initial (DFT) $\varepsilon_{n\textbf{k}_c}$ for any state $\textbf{k}_c = \textbf{k}_f$ inside the inner window. Moreover, the corresponding Bloch states

\begin{equation}\label{eq:Bloch_inside_inner}
\ket{\tilde{u}_{n\textbf{k}_c}} = \sum_{m \in \mathrm{dis}(\textbf{k}_c)} \ket{u_{m\textbf{k}_c}^\mathrm{sub}} U_{mn}(\textbf{k}_c)
\end{equation}
also coincide with the initial $\ket{u_{n\textbf{k}_c}}$ states inside the inner window.

The physical meaning of the resulting diagonalization matrix $U(\textbf{k}_f)$ is clearly visible from the definition of the Wannier-projected DOS (WDOS). For a set $\mathcal{W}$ of MLWFs it reads:

\begin{equation}\label{eq:WDOS_1}
N_{\mathcal{W}}(\varepsilon) = \sum_{n\textbf{k}} \bra{\tilde{u}_{n\textbf{k}}} \hat{\mathcal{P}}_{\mathcal{W}}(\textbf{k}) \ket{\tilde{u}_{n\textbf{k}}} \delta(\varepsilon_{n\textbf{k}} - \varepsilon),
\end{equation}

\begin{equation}\label{eq:WDOS_2}
\hat{\mathcal{P}}_{\mathcal{W}}(\textbf{k}) = \sum_{m \in \mathcal{W}} \ket{u_{m\textbf{k}}^\mathrm{sub}} \bra{u_{m\textbf{k}}^\mathrm{sub}}.
\end{equation}

Combining last three expressions, one obtains more explicit form for WDOS:

\begin{equation}\label{eq:WDOS_3}
N_{\mathcal{W}}(\varepsilon) = \sum_{n\textbf{k}} p_{n\textbf{k}}^{\mathcal{W}} \delta(\varepsilon_{n\textbf{k}} - \varepsilon), \;\;\; p_{n\textbf{k}}^{\mathcal{W}} = \sum_{m \in \mathcal{W}} |U_{mn}(\textbf{k})|^2.
\end{equation}
Thus, module-squared matrix elements of $U(\textbf{k}_f)$ serve as weights for extracting partial density of states, projected onto a chosen subset $\mathcal{W}$ of Wannier functions. For example, if one is interested in partial WDOS connected with more delocalized MLWFs of $s$-type:

\begin{equation}\label{eq:WDOS_s}
N_{s}(\varepsilon) = \sum_{n\textbf{k}} p_{n\textbf{k}}^{s} \delta(\varepsilon_{n\textbf{k}} - \varepsilon), \;\;\; p_{n\textbf{k}}^{s} = \sum_{m \in \mathcal{W}_s} |U_{mn}(\textbf{k})|^2,
\end{equation}
or, on the contrary, with more localized $d$-orbitals:

\begin{equation}\label{eq:WDOS_d}
N_{d}(\varepsilon) = \sum_{n\textbf{k}} p_{n\textbf{k}}^{d} \delta(\varepsilon_{n\textbf{k}} - \varepsilon), \;\;\; p_{n\textbf{k}}^{d} = \sum_{m \in \mathcal{W}_d} |U_{mn}(\textbf{k})|^2.
\end{equation}
If the total set of MLWFs consists only of two subsets $\mathcal{W}_s$ and $\mathcal{W}_d$, the full weight must be equal to unity for each state:

\begin{equation}\label{eq:ps_plus_pd}
p_{n\textbf{k}}^{s} + p_{n\textbf{k}}^{d} = 1,
\end{equation}
which is attributed to the unitary properties of the diagonalization matrix $U(\textbf{k})$.

\subsubsection{Electron-phonon coupling}

In a standard computational worflow, one usually starts from evaluatng EPC matrix elements on coarse \textbf{k}- and \textbf{q}-point grids in Bloch representation. There are two approaches frequently exploited, namely, method based on finite-differences evaluation, and the one built upon denisty functional perturbation theory (DFPT)~\cite{BaroniDFPT, giustino2017, DeformPot2021}. Independetly of the method used, EPC matrix element in Bloch representation reads (we use a different notation to distinguish these quantities from those obtained for dense grids after Wannier interpolation):
\begin{equation}\label{eq:EPC_Bloch_def}
\mathcal{G}_{mn}^{\nu}(\textbf{k},\textbf{q}) = \bra{u_{m\textbf{k}+\textbf{q}}} \Delta_{\textbf{q}\nu} v^\mathrm{KS} \ket{u_{n\textbf{k}}}_\mathrm{uc},
\end{equation}
where integral is carried out over a single unit cell of a crystal and $\Delta_{\textbf{q}\nu} v^\mathrm{KS}$ stands for the cell-periodic variation of the Kohn-Sham potential in the first order by ion collective displacements:
\begin{equation}
\Delta_{\textbf{q}\nu} v^\mathrm{KS} = \frac{1}{\sqrt{2 \omega_{\textbf{q}\nu}}} \sum_{\kappa \alpha \textbf{R}_p} e^{-i \textbf{q} (\textbf{r} - \textbf{R}_p)} \frac{e_{\kappa \alpha}^{\nu}(\textbf{q})}{\sqrt{M_\kappa}} \left. \frac{\partial V^\mathrm{KS}}{\partial \tau_{\kappa \alpha}} \right|_{\textbf{r} - \textbf{R}_p}.
\end{equation}
Here the sum extends over all the unit cells $\textbf{R}_p$ in a Wingner-Seitz supercell (of size $N_p$ unit cells) built based on atomic positions $\boldsymbol{\tau}_{\kappa p} = \textbf{R}_p + \boldsymbol{\tau}_\kappa$, $\kappa = 1, 2, ..., N_a$ with $N_a$ be the number of atoms in a unit cell; $M_\kappa$ denotes mass of the $\kappa$-th ion, $\omega_{\textbf{q}\nu}$ and $\textbf{e}_{\kappa}^{\nu}(\textbf{q})$ stand for eigenvalues and eigenmodes of phonon dynamical matrix; $\alpha = x, y$ or $z$ is the cartesian index of $\kappa$-th atom displacement induced by the corresponding phonon.

Following Giustino~\cite{giustino2007}, disentaglement matrices evaluated at the first stage as described in the previous subsection are used to construct another set of matrix elements by reducing dimensions into the optimal subspace and making the necessary rotation:

\begin{equation}\label{eq:G_after_disentangl}
\tilde{\mathcal{G}}_{mn}^{\nu}(\textbf{k}_c,\textbf{q}_c) = [\mathcal{U}^{\mathrm{dis}, \dagger}(\textbf{k}_c + \textbf{q}_c) \mathcal{G}^{\nu}(\textbf{k}_c,\textbf{q}_c) \mathcal{U}^{\mathrm{dis}}(\textbf{k}_c)]_{mn}.
\end{equation}
Here the part of matrix $\mathcal{G}^{\nu}(\textbf{k}_c,\textbf{q}_c)$ corresponding to disentaglement window is assumed; the left hand side matrix is of shape $N_\mathrm{sub} \times N_\mathrm{sub}$. Next, EPC matrix elements in Wannier representation are found by evaluating double Fourier series:
\begin{multline}\label{eq:G_in_Wannier}
g_{mn}^{\kappa \alpha}(\textbf{R}_e, \textbf{R}_p) = \frac{1}{N_c^e N_c^p}\sum_{\textbf{k}_c, \textbf{q}_c} e^{-i(\textbf{k}_c \textbf{R}_e + \textbf{q}_c \textbf{R}_p)} \\ \times \sum_{\nu} \tilde{\mathcal{G}}_{mn}^{\nu}(\textbf{k}_c,\textbf{q}_c) u^{-1}_{\kappa \alpha, \nu}(\textbf{q}_c),
\end{multline}
where $N_c^e$ and $N_c^p$ stand for coarse $\textbf{k}$- and $\textbf{q}$-grids' sizes, respectively, and $u^{-1}_{\kappa \alpha, \nu}(\textbf{q}) = \sqrt{2 M_\kappa \omega_{\textbf{q}\nu}} (e^\nu_{\kappa \alpha}(\textbf{q}))^{*}$. The inverse transformation yields the desired interpolated matrix elements:

\begin{multline}\label{eq:G_interpolated}
g_{mn}^{\nu}(\textbf{k}_f, \textbf{q}_f) = \sum_{\textbf{R}_e, \textbf{R}_p} e^{i(\textbf{k}_f \textbf{R}_e + \textbf{q}_f \textbf{R}_p)} \\ \times \sum_{\kappa \alpha} [U(\textbf{k}_f + \textbf{q}_f) g^{\kappa \alpha}(\textbf{R}_e, \textbf{R}_p) U^\dagger(\textbf{k}_f)]_{mn} u_{\kappa \alpha, \nu}(\textbf{q}_f),
\end{multline}
where $U^\dagger(\textbf{k}_f)$  and $U(\textbf{k}_f + \textbf{q}_f)$ are diagonalization matrices as defined in \eqref{eq.diagHam} and $u_{\kappa \alpha, \nu}(\textbf{q}) = 1/\sqrt{2 M_\kappa \omega_{\textbf{q}\nu}} e^\nu_{\kappa \alpha}(\textbf{q})$.

Most important quantity of EPC coupling in Wannier representation is attributed to its rapid decay as $|\textbf{R}_e|$ and $|\textbf{R}_p|$ grow.

\subsubsection{Construction of MLWFs in practice: initial projections and disentanglement strategies}

A crucial aspect of constructing MLWFs is the choice of an appropriate initial guess for the target orbitals. Achieving small Wannier spreads requires that the specific chemical and structural features of the material be taken into account, allowing the user to apply chemical intuition about bonding in the crystal of interest~\cite{oliphant2025bridging}. Most widely used method for providing these initial guesses is the \enquote{Wannier functions via projections} approach~\cite{MLWF2012}, in which one specifies both the center and the orbital character ($s$, $p$, $d$, or more complex, $sp^3$, $sp^3d^2$, etc) of each Wannier function as supported by \texttt{Wannier90} package~\cite{Wannier90_2008, Wannier90_2020}. The number of generated MLWFs equals the dimension of the optimal subspace $N_\mathrm{sub}$ of Bloch states selected for interpolation which may coincide with the total number of bands if no disentaglement is conducted. Consequently, it is common to employ non‑trivial projection sets, such as functions centred at interstitial sites rather than strictly at atomic positions. An appropriate selection of such projections accelerates the convergence of the Wannier spreads and improves the accuracy of the electron‑phonon matrix elements subsequently interpolated with \texttt{EPW}.

To construct a set of MLWFs that faithfully represent the hybrid $s$–$d$ complex in fcc palladium, we adopted the strategy described in the pioneering work of Souza, Marzari, and Vanderbilt (SMV) \cite{souza2001}. The authors considered two conceptually different tasks. The first, hereafter referred to as the \enquote{direct} approach, aims at isolating a broad $s$‑like band that significantly overlaps a $d$‑manifold. The authors employed their SMV algorithm to extract an optimal subspace of dimension $N_\mathrm{sub} = 5$ from an outer disentaglement energy window that also contains dispersive bands. Outer window subspaces of dimensions $6$ ($\mathcal{S}_6$) and $7$ ($\mathcal{S}_7$) were examined. By constructing the complement of the chosen subspace, they were able to isolate directly the wide $s$‑type band that intersects the $d$ bands. The present problem plays an important role for our research as a fully \textit{ab initio} treatment here enables the isolation of a $d$‑manifold for the subsequent calculation of its electronic properties. Simultaneously, the corresponding data for a broad $s$‑band become available. Nevertheless, this \enquote{direct} extraction of two highly disparate subspaces often suffers from a lack of physical rigor, an issue that will be discussed in later in more details.  

In contrast, the second task introduced by SMV aims at a more physically sound description of the subtle features of a typical transition metal electronic structure, explicitly accounting for the hybridization of states around the chemical potential; we will refer to this as the \enquote{hybridized} approach below. Interestingly, the authors demonstrated a clear advantage of working with the optimal subspace $\mathcal{S}_7$ compared to $\mathcal{S}_6$. In both cases the initial projections for the MLWFs consisted of five $d$‑orbitals centred on a Cu atom and $s$‑like orbitals placed at the tetrahedral interstitial sites of an fcc lattice. There was only a single $s$‑type orbital included in $\mathcal{S}_6$ construction, which broke the symmetry of a primitive fcc cell. To restore the full symmetry the authors introduced the second, equivalent $s$‑orbital, thereby forming the more symmetric subspace $\mathcal{S}_7$. As a result, two $s$-like orbitals were obtained located at positions $(1/4, 1/4, 1/4)$ and $(-1/4, -1/4, -1/4)$ in lattice vectors of a unit fcc cell. It should be emphasized that, in the \enquote{hybridized} approach, a sufficiently wide inner window is crucial for achieving highly accurate Wannier interpolation, whereas this window is irrelevant in the \enquote{direct} approach.

In this work we also carried out a detailed analysis of another common cubic crystal type lattice encountered in transition metals, namely the body‑centered cubic (bcc) structure. This choice was motivated primarily by the need for an additional benchmark of the transport properties results that are discussed later in the manuscript. Moreover, bcc crystals were not examined in the original SMV study, providing an additional argument for assessing the suitability of \enquote{direct} versus \enquote{hybridized} approaches. As a concrete test case we selected molybdenum.

\begin{figure}[htbp]
     \centering
     \includegraphics[width=0.45\textwidth]{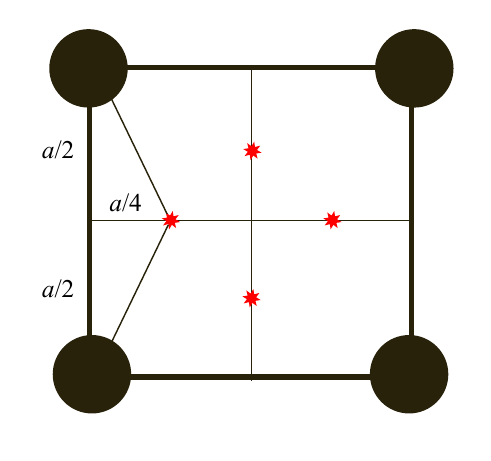}
     \caption{Locations of $s$-type MLWFs centers on a face of a conventional bcc cell. We employed an approach of constructing $\mathcal{S}_6$ optimal subspace with the 5 $d$-type MLWFs centered on an atom and the single $s$-type Wannier orbital which moves to one of four equivalent positions shown as red stars during spread functional minimization procedure.}
     \label{fig:bcc_wannier_centers}
 \end{figure}

When dealing with bcc metals we had to give special consideration to the choice of initial projections for constructing Wannier functions. Other authors commonly adopt the following nine function scheme with all orbitals centered on the atom: three $d$-orbitals of $t_{2g}$ symmetry and six $sp^3d^2$-orbitals. This method was compared to another set consisting of an $s$-orbital, three $p$-orbitals, and five $d$-orbitals, all atom-centered. It was shown that after spread-minimization procedure the latter set becomes the former one~\cite{Wang2006}. Our calculations support this. However, for the purposes of the present work this strategy is unsuitable, since it does not produce five well‑localized $d$‑type Wannier functions required for a clear projection and extraction of $d$‑state contribution $p_{n\textbf{k}}^{d}$ in \eqref{eq:WDOS_d}.

We tried to adopt an idea close to this of SMV for a bcc crystal lattice. We considered the case of symmetrized elementary bcc vectors cell basis. Again, we constructed 5 MLWFs, all atom-centered, and this time only a single $s$-type orbital per cell ($\mathcal{S}_1$ subspace). Regardless of the initial position of its center, minimization of the spread functional displaces the $s$-orbital to one of four symmetry-equivalent points on a square face of the conventional cubic cell, marked by red stars in Fig.~\ref{fig:bcc_wannier_centers}. Each of these points is equidistant from four atoms. In Fig.~\ref{fig:bcc_wannier_centers}, line segments are drawn explicitly to two of them, the corner atoms of the face; the remaining two sit at the body centers of the adjacent conventional cells that share this face. The segment connecting the latter pair is perpendicular to the plane of the figure and passes exactly through the center of the square.

\subsection{\textit{s}-\textit{d} scattering in Allen's formalism}

In this subsection we use Wannier formalism presented in the previous subsection to provide a generalized formulas in Allen's approach. Our objective is to explicitly account for the significant changes in orbital states of electrons before and after elementary scattering events, driven by strong hybridization in the vicinity of chemical potential. In the first step, we rewrite the elementary transition probability defined in \eqref{eq:Pkk_for_elph} as follows

\begin{widetext}

\begin{multline}\label{eq:Pkk_sd}
P_{n\textbf{k}, s}^{m\textbf{k}+\textbf{q}, d} = \frac{2\pi}{\hbar} \sum_{\textbf{q}\nu} |g_{mn}^{\nu}(\textbf{k},\textbf{q})|^2 p_{n\textbf{k}}^{s}f(\varepsilon_{n\textbf{k}}) p_{m\textbf{k}+\textbf{q}}^{d} (1 - f(\varepsilon_{m\textbf{k}+\textbf{q}})) \\ \times [(n(\omega_{\textbf{q}\nu}) + 1)\delta(\varepsilon_{n\textbf{k}} - \varepsilon_{m\textbf{k}+\textbf{q}} - \omega_{\textbf{q}\nu}) + n(\omega_{\textbf{q}\nu})\delta(\varepsilon_{n\textbf{k}} - \varepsilon_{m\textbf{k}+\textbf{q}} + \omega_{\textbf{q}\nu})]
\end{multline}
for the special case of $s$-$d$ contribution. The other three contributions are easily written accordingly. Here $p_{n\textbf{k}}^{s}$ is given in \eqref{eq:WDOS_s} and $p_{m\textbf{k}+\textbf{q}}^{d} = 1 - p_{m\textbf{k}+\textbf{q}}^{s}$. The form given in \eqref{eq:Pkk_sd} corresponds to the product rule for the probabilities of independent events. It follows that the expression for elementary probability of electron scattering in the $s$-channel is

\begin{multline}\label{eq:Pkk_s}
P_{n\textbf{k}, s}^{m\textbf{k}+\textbf{q}} = P_{n\textbf{k}, s}^{m\textbf{k}+\textbf{q}, s} + P_{n\textbf{k}, s}^{m\textbf{k}+\textbf{q}, d} =  \frac{2\pi}{\hbar} \sum_{\textbf{q}\nu} |g_{mn}^{\nu}(\textbf{k},\textbf{q})|^2 p_{n\textbf{k}}^{s}f(\varepsilon_{n\textbf{k}}) (1 - f(\varepsilon_{m\textbf{k}+\textbf{q}})) \\ \times [(n(\omega_{\textbf{q}\nu}) + 1)\delta(\varepsilon_{n\textbf{k}} - \varepsilon_{m\textbf{k}+\textbf{q}} - \omega_{\textbf{q}\nu}) + n(\omega_{\textbf{q}\nu})\delta(\varepsilon_{n\textbf{k}} - \varepsilon_{m\textbf{k}+\textbf{q}} + \omega_{\textbf{q}\nu})].
\end{multline}
Similar expression holds for the $d$-channel.

Thus, to account for the strong hybridization between an $s$-band and a $d$-complex in a transition metal, we replace the unit weight of each Bloch state with its actual fractional $s$-orbital weight when considering the $s$-channel. This weighting factor must be applied to both the numerator (transition rates) and the denominator (normalization sum) of the expression for TSF to ensure statistical consistency across the hybridized state space. So we accordingly rewrite the transport DOS ($x$-component is assumed and $\alpha$ label is ommited) first introduced in \eqref{eq:TrDOS}:

\begin{equation}\label{eq:TrDOS_s}
N_s(\varepsilon)v^2(\varepsilon) =  \sum_{n\textbf{k}} p_{n\textbf{k}}^{s} v_{n\textbf{k}}^2 \delta(\varepsilon_{n\textbf{k}} - \varepsilon)
\end{equation}

\begin{equation}\label{eq:TrDOS_d}
N_d(\varepsilon)v^2(\varepsilon) =  \sum_{n\textbf{k}} p_{n\textbf{k}}^{d} v_{n\textbf{k}}^2 \delta(\varepsilon_{n\textbf{k}} - \varepsilon)
\end{equation}

Below we provide the explicit formulas for an $s$-$d$ and $s$-$s$ contributions to electrical resistivity in the $s$-channel, the $d$-channel contributions are written accordingly:

\begin{equation}\label{eq:rho_from_Q_sd}
\rho_{sd}(T) = \frac{1}{2e^2} Q_{x0,x0}^{sd}(T),
\end{equation}

\begin{multline}\label{eq:ord0ExactFormulas_sd}
Q_{x0,x0}^{sd} =
\dfrac{2 \pi V}{\hbar T}
\int d\Omega
\int  d\varepsilon \dfrac{I_0^{sd}(\varepsilon,\Omega)}{N_s(\varepsilon)v^2(\varepsilon)}, \\
I_0^{sd}(x,x,\varepsilon,\Omega) = 
\alpha^2_{\mathrm{tr}, sd} F(1,1,\varepsilon,\Omega)
\int d\varepsilon^\prime f(\varepsilon, T) (1 - f(\varepsilon^\prime, T)) \\ \times
[(n(\Omega, T) + 1)\delta(\varepsilon - \varepsilon^\prime - \Omega) + 
n(\Omega, T)\delta(\varepsilon - \varepsilon^\prime + \Omega)],
\end{multline}

\begin{multline}\label{eq:TSF_sd}
\alpha^2_{\mathrm{tr},sd}F(1, 1, \varepsilon, \Omega) = \frac{1}{2N_{s}(\varepsilon)v^2(\varepsilon)} \sum_{\textbf{q}\nu} \delta(\Omega - \omega_{\textbf{q}\nu}) \sum_{mn,\textbf{k}} |g^\nu_{mn}(\textbf{k},\textbf{q})|^2
    (v_{n\textbf{k}} - v_{m\textbf{k} + \textbf{q}})^2 p_{n\textbf{k}}^{s} \delta(\varepsilon_{n\textbf{k}} - \varepsilon) p_{m\textbf{\textbf{k}+\textbf{q}}}^{d} \delta(\varepsilon_{m\textbf{k}+\textbf{q}} - \varepsilon),
\end{multline}

\begin{equation}\label{eq:rho_from_Q_ss}
\rho_{ss}(T) = \frac{1}{2e^2} Q_{x0,x0}^{ss}(T),
\end{equation}

\begin{multline}\label{eq:ord0ExactFormulas_ss}
Q_{x0,x0}^{ss} =
\dfrac{2 \pi V}{\hbar T}
\int d\Omega
\int  d\varepsilon \dfrac{I_0^{ss}(\varepsilon,\Omega)}{N_s(\varepsilon)v^2(\varepsilon)}, \\
I_0^{ss}(x,x,\varepsilon,\Omega) = 
\alpha^2_{\mathrm{tr}, ss} F(1,1,\varepsilon,\Omega)
\int d\varepsilon^\prime f(\varepsilon, T) (1 - f(\varepsilon^\prime, T)) \\ \times
[(n(\Omega, T) + 1)\delta(\varepsilon - \varepsilon^\prime - \Omega) + 
n(\Omega, T)\delta(\varepsilon - \varepsilon^\prime + \Omega)],
\end{multline}

\begin{multline}\label{eq:TSF_ss}
\alpha^2_{\mathrm{tr},ss}F(1, 1, \varepsilon, \Omega) = \frac{1}{2N_{s}(\varepsilon)v^2(\varepsilon)} \sum_{\textbf{q}\nu} \delta(\Omega - \omega_{\textbf{q}\nu}) \sum_{mn,\textbf{k}} |g^\nu_{mn}(\textbf{k},\textbf{q})|^2
    (v_{n\textbf{k}} - v_{m\textbf{k} + \textbf{q}})^2 p_{n\textbf{k}}^{s} \delta(\varepsilon_{n\textbf{k}} - \varepsilon) p_{m\textbf{\textbf{k}+\textbf{q}}}^{s} \delta(\varepsilon_{m\textbf{k}+\textbf{q}} - \varepsilon),
\end{multline}

\end{widetext}

Another question which arises when extracting different contributions into electrical resistivity is how all of them are connected to the full resistivity $\rho$, obtained by standard Allen's approach in \eqref{eq:rho_from_Q} and \eqref{eq:ord0ExactFormulas}. It was mentioned by different authors that resistivities from $s$- and $d$-channels must be added as parallel conductors~\cite{Weger1985, Irkhin2021}:

\begin{equation}\label{eq:rho_parallel}
\frac{1}{\rho_{||}} = \frac{1}{\rho_{s}} + \frac{1}{\rho_{d}},
\end{equation}
whereas the resistivities in separate channels must be added in series:

\begin{equation}\label{eq:rho_s_chan}
\rho_s = \rho_{ss} + \rho_{sd}
\end{equation}
and

\begin{equation}\label{eq:rho_d_chan}
\rho_d = \rho_{ds} + \rho_{dd}.
\end{equation}

The decomposition of the total resistivity into four parts proposed above implies several expected properties, which we now list and then verify in the actual calculations below. First, $\rho_{dd}$ part should be close to the full resistivity $\rho$ in standard approach \eqref{eq:rho_from_Q} due to significant majority of electrons that are in a nearly pure $d$-states near the chemical potential in transition metals. At the same time, $\rho_{ds}$ part should be much less than that of $\rho_{dd}$. Next, $\rho_{sd}$ contribution is assumed to be much larger than all the other three for the parallel conductors expression \eqref{eq:rho_parallel} to be close to the full standard resistivity in \eqref{eq:rho_from_Q}.

In a similar way one can consider the parallel addition of channels for electronic thermal conductivity. In this case the conductivities in the $s$- and $d$-channels must be summed up:

\begin{equation}\label{eq:kappa_parallel}
\kappa_{||}^{e} = \kappa_{s}^{e} + \kappa_{d}^{e},
\end{equation}
and conductivities in separate channels must be added in parallel:

\begin{equation}\label{eq:kappa_both_chan}
\frac{1}{\kappa_{s}^{e}} = \frac{1}{\kappa_{ss}^{e}} +\frac{1}{\kappa_{sd}^{e}}, \;\;\; \frac{1}{\kappa_{d}^{e}} = \frac{1}{\kappa_{ds}^{e}} +\frac{1}{\kappa_{dd}^{e}}.
\end{equation}
The expressions for each contribution are the generalized versions of \eqref{eq:kappa_from_Q}, \eqref{eq:ord1ExactFormulas} and \eqref{eq:ord2ExactFormulas} in accordance with those for resistivities above.


\subsection{Computational details}\label{subsec.comput}

\begin{figure*}[htbp]
    \centering
    \begin{subfigure}[b]{0.28\textwidth}
        \centering
        \includegraphics[width=\textwidth]{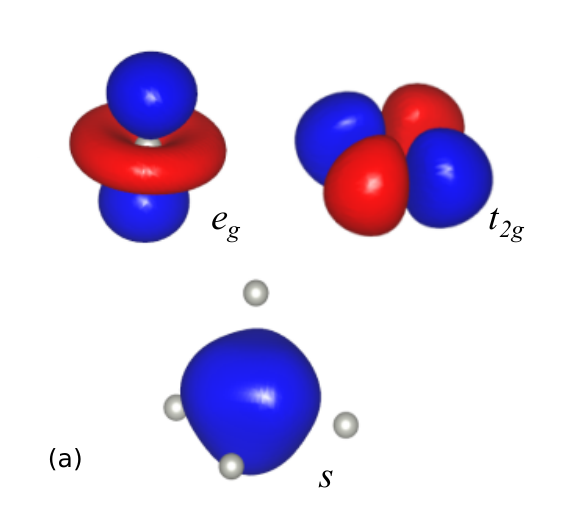}
    \end{subfigure}
    \hfill%
    \begin{subfigure}[b]{0.34\textwidth}
        \centering
        \includegraphics[width=\textwidth]{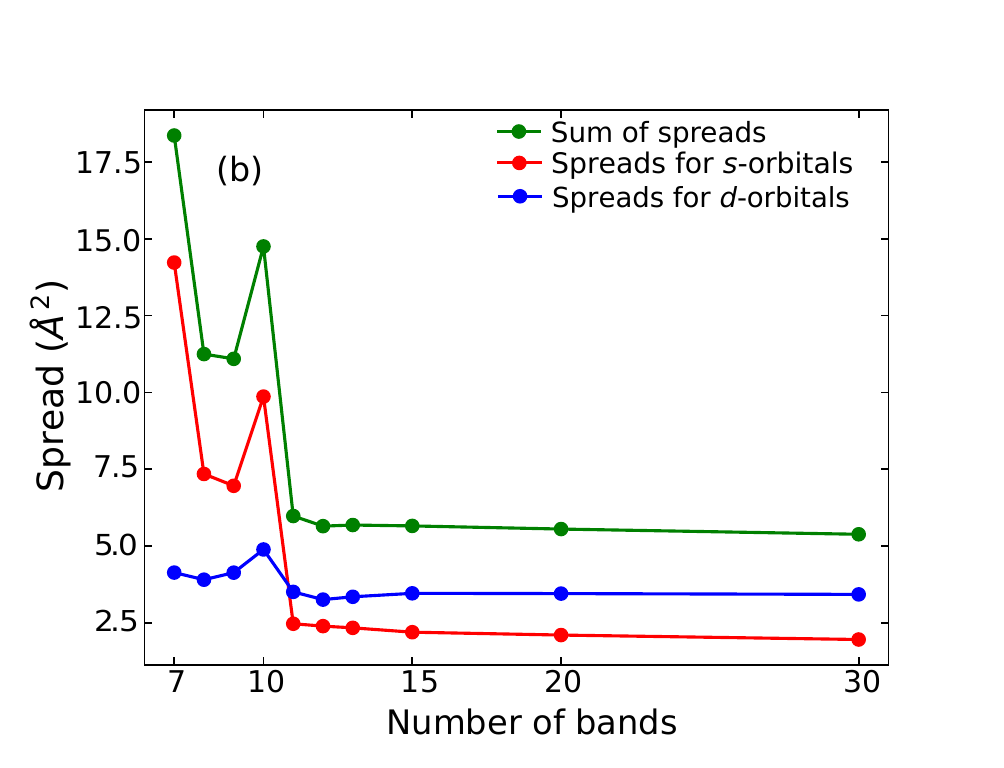}
    \end{subfigure}
    \hfill%
    \begin{subfigure}[b]{0.34\textwidth}
        \centering
        \includegraphics[width=\textwidth]{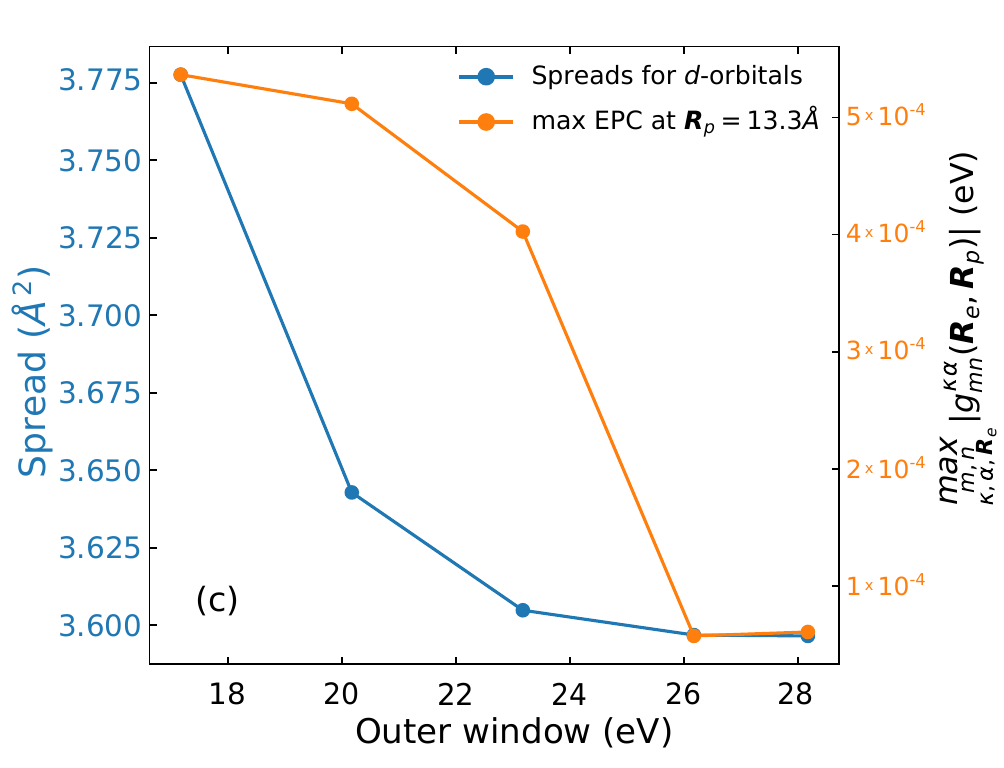}
    \end{subfigure}
    \caption{Construction of MLWFs for fcc Pd following ideas of SMV. (a) Isosurafces of obtained atom-centered $e_g$, $t_{2g}$ orbitals of $d$-type and one $s$-type orbital in a tetrahedral interstitial site of fcc unit cell obtained in optimal $\mathcal{S}_7$ subspace. Red and blue colors represent the positive and negative amplitudes of the real-valued MLWFs, respectively. (b) Convergence of the spreads as a function of the number of bands used in the construction of the MLWF basis for $\mathcal{S}_7$. (c) Convergence of the total $d$-orbital spread and the maximum EPC matrix element at large distances from the home cell as a function of the outer window width when constructing $\mathcal{S}_5$ optimal subspace.}
    \label{fig:wannier_convergence_Pd}
\end{figure*}

\begin{figure}[b]
     \centering
     \includegraphics[width=0.45\textwidth]{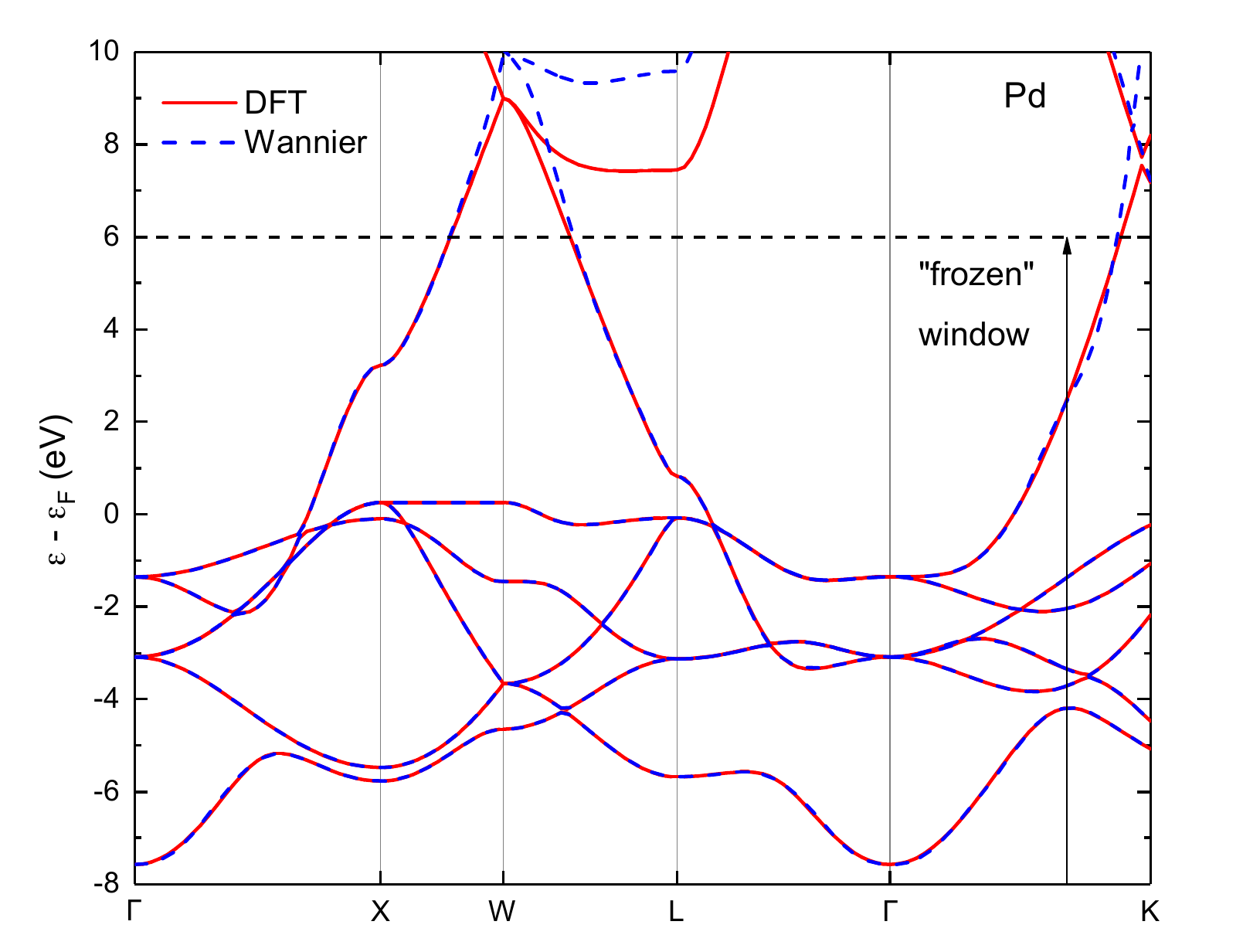}
     \caption{Comaprison of DFT and Wannier-interpolated band structure in Pd. Blue line denotes the DFT results, dashed red line is for Wannier interpolation.}
     \label{fig:Pd_bands_DFT_vs_W}
 \end{figure}

\begin{figure*}[htbp]
     \centering
     \includegraphics[width=1.0\textwidth]{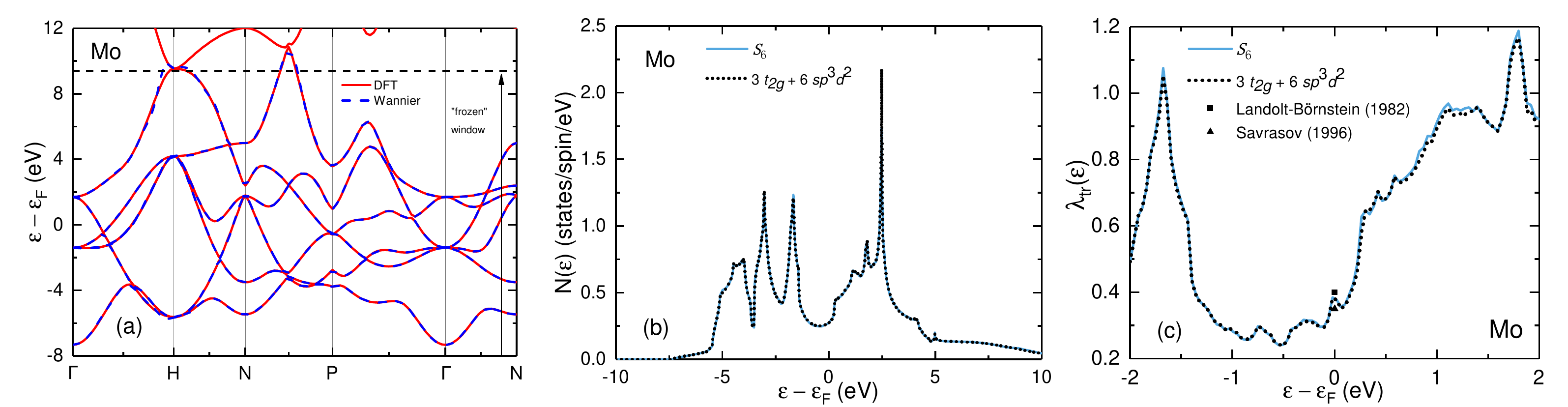}
     \caption{Construction of MLWFs for bcc Mo. (a) Comaprison of DFT and Wannier-interpolated band structure in Mo. Blue line denotes the DFT results, dashed red line is for Wannier interpolation. The chosen $\mathcal{S}_6$ subspace perfectly describes the disentanglement of upper $s$-band with the $d$-complex. (b) Comparison of electron DOS between $sp^3d^2$ initial projections and $\mathcal{S}_6$ approach applied in this work. (c) Same comparison for transport EPC coupling $\lambda_\mathrm{tr}$ in a range of electron energies. Also the value inferred from experimental data by \cite{landolt1978numerical} is shown as black square at Fermi energy, along with the numerical results by \cite{Savrasov1996} shown as black triangle.}
     \label{fig:wannier_convergence_bcc}
 \end{figure*}

All first‑principles calculations in this work were carried out within \texttt{Quantum Espresso} software suite~\cite{QE2017}. The computational workflow can be described as three successive stages. At the first stage, electronic structure and lattice dynamics are obtained on relatively coarse $\textbf{k}$‑point meshes in reciprocal space using density functional theory (DFT) and density functional perturbation theory (DFPT), respectively. The second stage employs \texttt{EPW} (Electron‑Phonon-Wannier) module~\cite{EPW2016, EPW2023} to generate the Hamiltonian matrix elements and the electron‑phonon interaction vertices in Wannier representation, as described above. At the final stage, we use our \texttt{SKiES} code~\cite{GALTSOV2025} to solve Boltzmann kinetic equation via Allen’s method including the generalisation that enables a separate treatment of distinct electronic contributions. Technical details of each stage are provided in the corresponding subsections below, with particular emphasis on the construction of the MLWFs at the second stage.

\subsubsection{DFT and DFPT calculations}\label{subsubsec:qe_part}

All calculations were performed with pseudopotential within the projector‑augmented‑wave (PAW) framework~\cite{PAW1994}. For both palladium and molybdenum we employed local-density approximation (LDA) parametrisation in the form of Perdew‑Zunger for exchange-correlation functional~\cite{LDA_PZ}. We started with a full structural relaxation for each investigated metal at $300$~K; the resulting lattice constants for Pd and Mo deviate from the corresponding experimental values by $1$\%, which is within the typical accuracy of first‑principles methods. The self‑consistent electronic‑structure calculations were carried out on a $20 \times 20 \times 20$ $\textbf{k}$‑point mesh with an energy‑convergence threshold of $10^{-12}$~Ry.  

Phonon properties were obtained using density‑functional perturbation theory (DFPT) on several comparatively coarse Monkhorst‑Pack $\textbf{q}$‑point grids, namely $6 \times 6 \times 6$ and $8 \times 8 \times 8$. We performed these additional convergence checks because the choice of the $\textbf{q}_c$‑grid determines the quality of the resulting phonon dispersions and the electron‑phonon matrix elements later interpolated with \texttt{EPW}. In all phonon calculations the self‑consistency threshold was tightened to $10^{-17}$~Ry to ensure an accurate description of phonon branches, especially in the vicinity of the $\Gamma$-point.

Subsequently, a non-self‑consistent calculation of electronic structure was carried out on coarse $\textbf{k}_c$-grids of sizes $8 \times 8 \times 8$ for Pd (the same as $\textbf{q}_c$-grid for Pd) and $12 \times 12 \times 12$ for Mo ($\textbf{q}_c$-grid of sizes $6 \times 6 \times 6$ was exploited for Mo as phonon dispersions were finally described better than for $8 \times 8 \times 8$). The number of electronic bands at this stage is a crucial parameter for the subsequent Wannier interpolation; the criteria for selecting this number are discussed in the following subsection. Throughout all DFT calculations a sufficiently large plane‑wave kinetic‑energy cutoff of $90$~Ry was employed.

\subsubsection{\texttt{EPW} calculations}\label{subsubsec:epw_part}

Pd is an fcc metal, so the original ideas of SMV are applicable unchanged for it. Fig.~\ref{fig:wannier_convergence_Pd}(a) shows several MLWFs obtained while constructing subspace $\mathcal{S}_7$ for Pd. Our tests show that the total spread of all $7$ Wannier functions are almost identical for coarse grids $6 \times 6 \times 6$ and $8 \times 8 \times 8$. In Fig.~\ref{fig:wannier_convergence_Pd}(b) we provide our convergence tests for sums of spreads as the number of bands involved in the outer window construction is increased. The width of the \enquote{frozen} window was set to $14$~eV to correctly describe both $s$-$d$ complex and the upper wide $s$-band. Results for separate subsets of $s$-orbitals as well as $d$-orbitals and sums of all obtained MLWFs are shown. The comparison between the DFT bands and those from Wannier interpolation for Pd is shown in Fig.~\ref{fig:Pd_bands_DFT_vs_W}. Despite small discrepancies above the frozen energy window in the disentanglement procedure, the perfect agreement is reached.

Within the \enquote{direct} approach we extracted an optimal subspace $\mathcal{S}_5$ that captures the $d$‑complex in Pd inside a larger $\mathcal{S}_7$ space. The convergence of the total Wannier spread of the five $d$‑type MLWFs was examined as a function of the outer window width; the results are shown by the blue curve in the Fig.~\ref{fig:wannier_convergence_Pd}(c). Convergence is already achieved for an outer window of close to $29$~eV (measured from the lowest eigenvalue at the $\Gamma$ point), which is the upper bound of $\mathcal{S}_7$. This width ensures that exactly seven bands (possibly degenerate) are present at each $\textbf{k}$‑point of the coarse mesh. The orange curve in Fig.~\ref{fig:wannier_convergence_Pd}(c) displays the corresponding maximum value of the EPC matrix element

$$\max_{mn\kappa\alpha \textbf{R}_e} |g_{mn}^{\kappa\alpha}(\textbf{R}_e, \textbf{R}_p = 13.3 \AA)|$$
far from the home cell in the Wannier representation, indicating a desired robust localization of EPC.


The comparison between the resulting Wannier-interpolated band structure from $\mathcal{S}_6$ construction and the reference DFT band structure for Mo is presented in Fig.~\ref{fig:wannier_convergence_bcc}(a). Fig.~\ref{fig:wannier_convergence_bcc}(b,c) illustrates the comparison of EPC properties for two distinct initial projection choices for MLWFs. Fig.~\ref{fig:wannier_convergence_bcc}(b) shows our results for electron DOS and (c) for EPC coupling function $\lambda_\mathrm{tr}(\varepsilon) = 2 \int_{0}^{\infty} d\Omega/\Omega \;\; \alpha_\mathrm{tr}^2F(1, 1, \varepsilon, \Omega)$. The first is for more popular six $sp^3d^2 +$ three $t_{2g}$ approach; the second is for proposed five $d$-orbitals centered on atom and one $s$-orbital ($\mathcal{S}_6$). It can be seen that employing the latter scheme gives the similar trend of the corresponding properties for wide ranges of electron energies around Fermi energy as the more common approach. We additionally provide in Fig.~\ref{fig:wannier_convergence_bcc}(c) results for $\lambda_\mathrm{tr}(\varepsilon_F)$ obtained based on experimental results for the electrical–resistivity data \cite{landolt1978numerical} and from \textit{ab initio} calculations by Savrasovs' \cite{Savrasov1996}. In all calculations, an outer disentanglement energy window of $95$~eV, encompassing a total of $30$ bands and ensuring well-converged results, was employed; the width of the inner window was set to $18$~eV. An excelent agreement here in the whole inner window additionally justifies our choice.

\subsubsection{Transport properties calculations}\label{subsubsec:skies_part}

All calculations for electrical resistivity and thermal electron conductivity in this work were made in the framework of our \texttt{SKiES} code \cite{GALTSOV2025}. The developed code relies on preliminary calculations performed using \texttt{EPW} package, which requires a set of files containing the Hamiltonian \eqref{eq.HamWan2Bloch}, electron velocities, dynamical phonon matrix, and electron-phonon coupling matrix elements \eqref{eq:G_in_Wannier} in Wannier basis. Various densities of states \eqref{eq:DOS}, \eqref{eq:TrDOS}, \eqref{eq:WDOS_3}  and Allen's transport spectral function \eqref{eq:A2F_pp}, \eqref{eq:A2F_mm} were computed using tetrahedron method~\cite{allen1983tetrahedron, lambin1984computation, BlochlTetra}. This approach provides a more accurate description of the fine features of these quantities compared to the widely used Gaussian smearing method (which is commonly employed for similar tasks within \texttt{EPW}).

We have thoroughly investigated the convergence of the transport spectral function with respect to the dense Brillouin zone grid size, employing grids up to $100 \times 100 \times 100$ for $\textbf{k}$-points and $25 \times 25 \times 25$ \textbf{q}-points. Such dense sampling is crucial for the metals considered in this study due to the complex topology of their Fermi surfaces.

To support calculations in the framework of proposed \enquote{hybridized} Allen's method, we have implemented the calculation of various partial densities of states \eqref{eq:WDOS_s}, \eqref{eq:WDOS_d} and orbital-resolved spectral functions \eqref{eq:TSF_sd}, \eqref{eq:TSF_ss}, obtained by projecting onto a user-specified set of Wannier functions taken from preliminary \texttt{EPW} calculations.

When integrating over electron energy range in general Allen's formulas \eqref{eq:ord0ExactFormulas}-\eqref{eq:ord2ExactFormulas} and \eqref{eq:ord0ExactFormulas_sd}-\eqref{eq:ord0ExactFormulas_ss} we additionally checked convergence with respect to the bounds around chemical potential. The final ranges of $[-0.8, 0.8]$~eV for Pd and $[-1.2, 1.2]$~eV for Mo were used in all calculations.

\section{Results and discussions}\label{sec:results}

\subsection{Pd}\label{subsec:Pd}

\begin{figure}[htbp]
    \centering
    \begin{subfigure}[b]{0.4\textwidth}
        \centering
        \includegraphics[width=\textwidth]{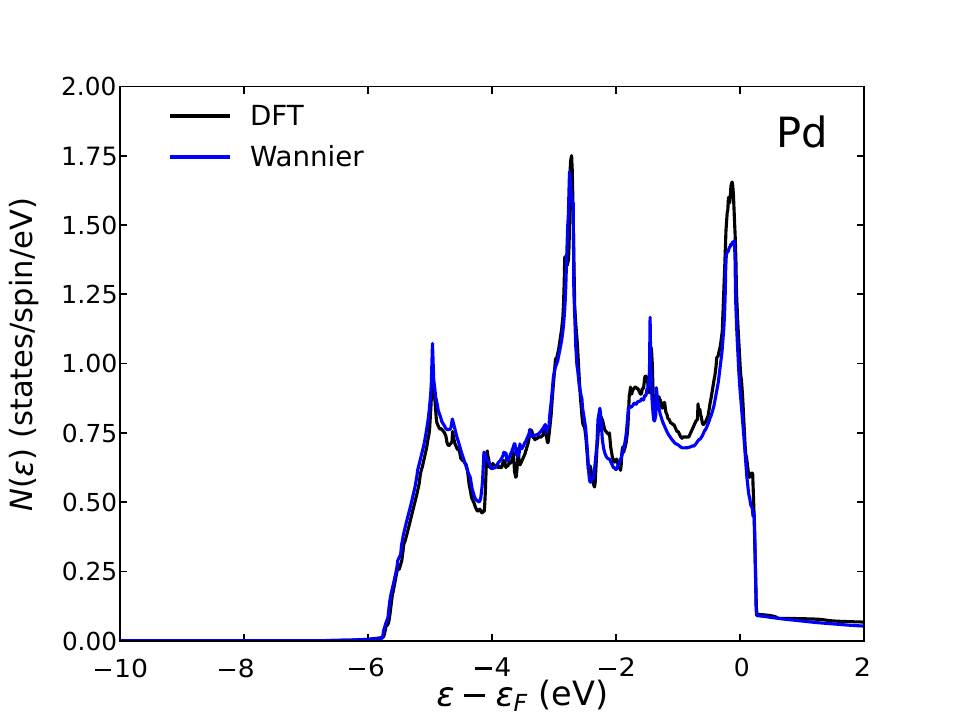}
    \end{subfigure}
    \vfill%
    \begin{subfigure}[b]{0.4\textwidth}
        \centering
        \includegraphics[width=\textwidth]{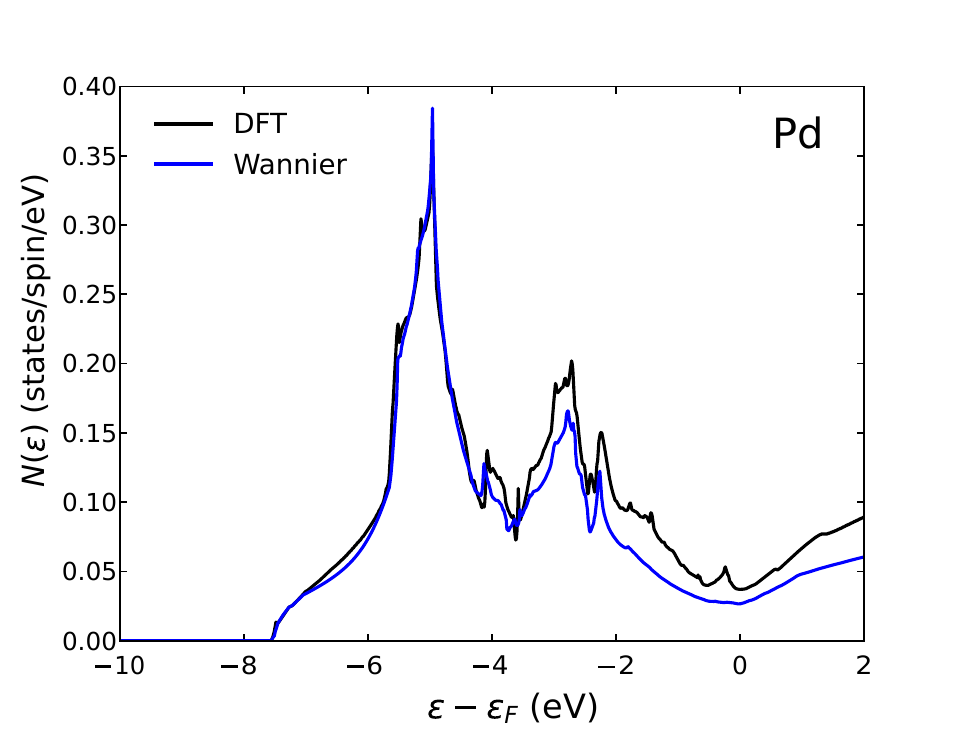}
    \end{subfigure}
    \caption{Partial DOS of Pd from direct DFT projections and Wannier interpolation. (a)~$d$ states; (b)~$s$ states. In the DFT data the $s$ panel shows the combined $s+p$ contribution, whereas the Wannier curve is the sum of two $s$-type projections ($\mathcal{S}_2$ subspace).}
    \label{fig:DOS_hybrid_DFT_vs_W}
\end{figure}

Figure~\ref{fig:DOS_hybrid_DFT_vs_W} compares the partial electronic densities of states of Pd obtained from direct DFT projections and from Wannier interpolation. 
The $d$-electron DOS Fig.~\ref{fig:DOS_hybrid_DFT_vs_W}(a) is virtually identical in the two approaches. For the $s$ states Fig.~\ref{fig:DOS_hybrid_DFT_vs_W}(b), the DFT curve contains the summed $s$ and $p$ contributions, whereas the Wannier result represents the projection onto the two $s$-type functions of the $\mathcal{S}_2$ subspace. Agreement remains over the whole energy window shown. The deviation in the region of $[-0.5, 0.5]$~eV around the Fermi level is only about $0.01$~electron/spin/eV which is insignificant. At the same time the characteristic minumum close to $\varepsilon_F$ is reproduced correctly.

\begin{figure*}[htbp]
     \centering
     \includegraphics[width=1.0\textwidth]{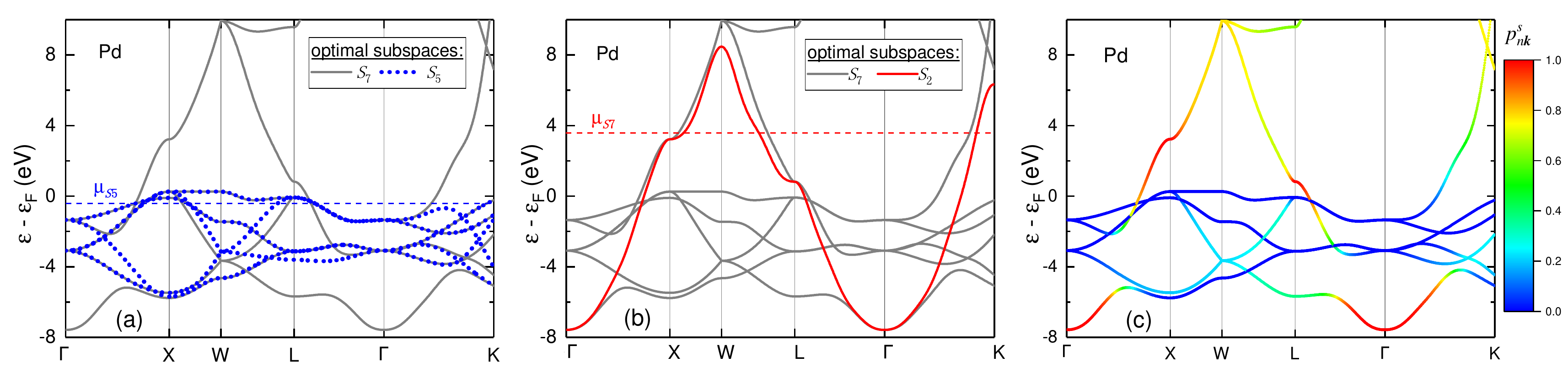}
     \caption{Band structure of Pd. (a)~Gray: full $\mathcal{S}_7$ subspace; blue dotted: $d$-complex from the optimal $\mathcal{S}_5$ subspace. (b)~Red: complementary $\mathcal{S}_2$ subspace. (c)~Same $\mathcal{S}_7$ bands as in (a), colored by the Wannier-projected $s$-state weight in $\mathcal{S}_2$ (Eq.~\eqref{eq:WDOS_s}).}
     \label{fig:direct_bands_Pd}
\end{figure*}

The $\mathcal{S}_5$ band structure is compared with the full $\mathcal{S}_7$ result in Fig.~\ref{fig:direct_bands_Pd}(a). 
Both dispersions coincide at the Fermi level, while the largest deviations appear near the W point and along the $\Gamma$--X and W--L--$\Gamma$--K segments roughly 3~eV below $\varepsilon_F$. The complementary $\mathcal{S}_2$ subspace Fig.~\ref{fig:direct_bands_Pd}(b) produces the expected broad, parabolic $s$-band. It matches the lowest $\mathcal{S}_7$ band at the $\Gamma$ point, where $s$--$d$ hybridization is negligible, and crosses the $d$ complex exactly along those high-symmetry paths that show the largest deviations in panel~(a).

Figure~\ref{fig:direct_bands_Pd}(c) recasts the $\mathcal{S}_7$ bands in the \enquote{hybridized} picture, coloring each state by its Wannier-projected $s$-character within $\mathcal{S}_2$ Eq.~\eqref{eq:WDOS_s}. 
Red denotes a dominant $s$ weight and deep blue marks pure $d$ states. 
The color variation is smooth, and the bands intertwine naturally, without the unphysical crossings inevitable in the direct construction. 
Regions where the $\mathcal{S}_2$ band comes closest to the $\mathcal{S}_7$ manifold in panel~(b)—near the bottom at $\Gamma$, just above $\varepsilon_F$ at L, and around X at around 3~eV above $\varepsilon_F$—coincide with $s$ fractions approaching unity. 
Conversely, the $\Gamma$--K segment above $\varepsilon_F$ retains a strong admixture of both characters, confirming that this part of the wide band cannot be regarded as purely delocalized.

As expected, the \enquote{direct} and \enquote{hybridized} approaches do not describe the same electron-phonon system. 
The reconstructed states in the optimal subspaces $\mathcal{S}_5$ and $\mathcal{S}_2$ correspond to a model in which the $d$- and $s$-like species are decoupled, whereas in $\mathcal{S}_7$ both share a single Fermi level (the zero of energy in all panels of Fig.~\ref{fig:direct_bands_Pd}), reflecting global equilibrium. 
Consequently, the chemical potentials in $\mathcal{S}_5$ and $\mathcal{S}_2$ must shift away from zero to preserve the proper electron counts:
\begin{equation}\label{eq:chem_pot_S5}
\int_{-\infty}^{\infty} d\varepsilon \frac{N_{\mathcal{S}_5}(\varepsilon)}{\exp \left( \frac{\varepsilon - \mu_{\mathcal{S}_5}}{T} \right) + 1} = N_d,
\end{equation}
\begin{equation}\label{eq:chem_pot_S2}
\int_{-\infty}^{\infty} d\varepsilon \frac{N_{\mathcal{S}_2}(\varepsilon)}{\exp \left( \frac{\varepsilon - \mu_{\mathcal{S}_2}}{T} \right) + 1} = N_s,
\end{equation}
with $N_s=1.5$ and $N_d=8.5$ electrons, summing to the ten valence electrons of the pseudopotential used. 
These occupations are obtained by integrating the partial DOS below $\varepsilon_F$ in the hybridized picture.

\begin{figure}[b]
     \centering
     \includegraphics[width=0.5\textwidth]{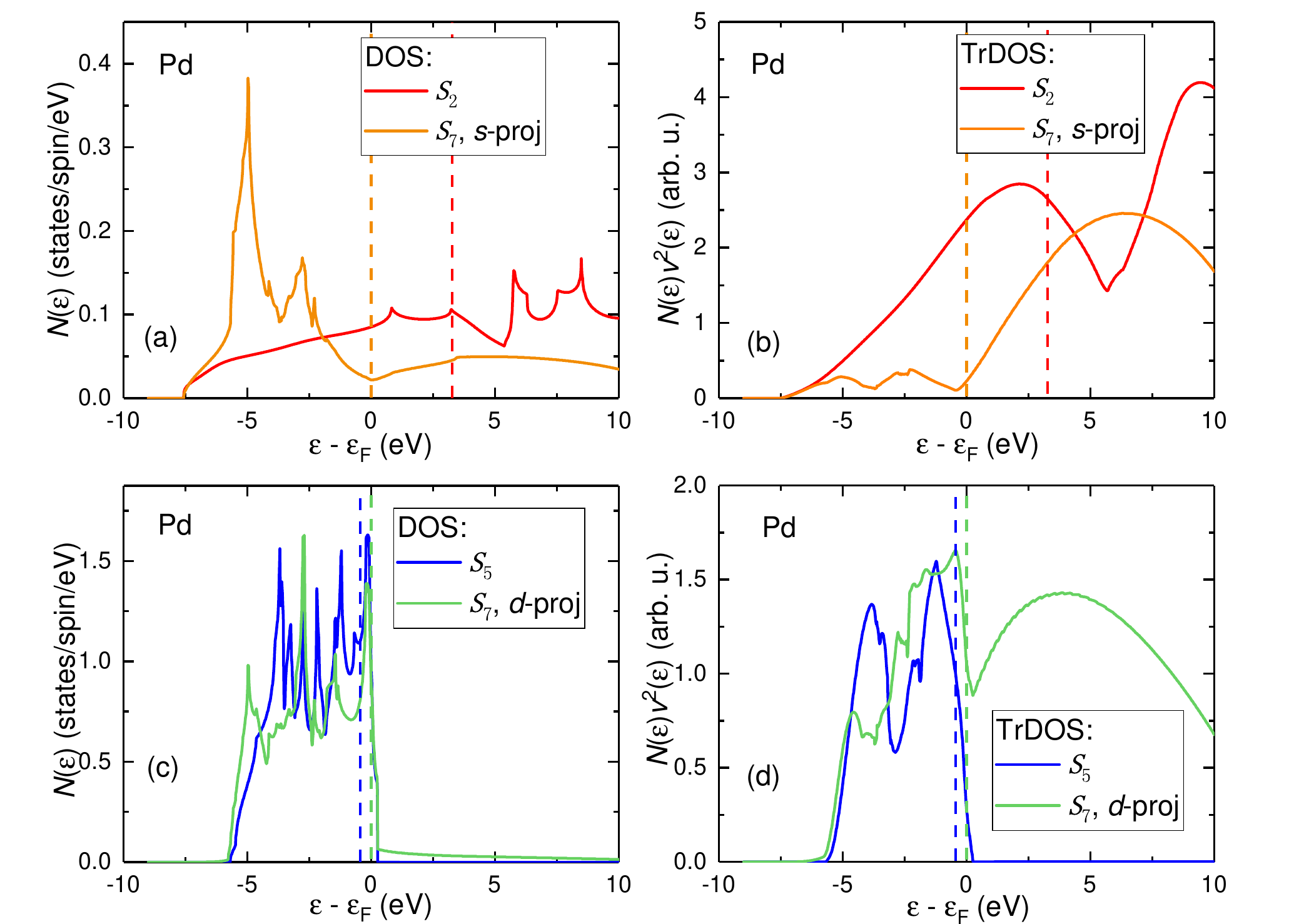}
     \caption{Densities of states in the direct (optimal subspaces) and hybridized (Wannier-projected) approaches for fcc Pd. (a)~DOS and (b)~TrDOS for $\mathcal{S}_2$ (red) versus the projected $s$ states in $\mathcal{S}_7$ (orange). (c)~DOS and (d)~TrDOS for $\mathcal{S}_5$ (blue) versus the projected $d$ states in $\mathcal{S}_7$ (green).}
     \label{fig:DOS_Pd_direct_vs_hybrid}
\end{figure}

The resulting densities of states are compared in Fig.~\ref{fig:DOS_Pd_direct_vs_hybrid}. 
For the $s$ states in panels~(a,b), the differences are pronounced: the $\mathcal{S}_7$ projection shows a peak roughly 5~eV below $\varepsilon_F$ and a deep minimum near the Fermi level, whereas the direct $\mathcal{S}_2$ result stays lower and retains a nearly free-electron-like square-root shape. 
The distinction is amplified in the transport DOS Fig.~\ref{fig:DOS_Pd_direct_vs_hybrid}(b): the projected $\mathcal{S}_7$ TrDOS is roughly an order of magnitude smaller than the direct $\mathcal{S}_2$ value. 
This reflects the band curvatures in Fig.~\ref{fig:direct_bands_Pd}(b): the broad parabolic $\mathcal{S}_2$ band demonstrates larger electronic velocity $v_{n\mathbf{k}}=\partial\varepsilon_{n\mathbf{k}}/\partial\mathbf{k}$ at the Fermi level than the flat $\mathcal{S}_7$ bands. 
For the $d$ states in panels~(c,d), the DOS curves are practically identical near $\varepsilon_F$, both displaying the characteristic narrow $d$-peak. 
In the TrDOS, however, the $\mathcal{S}_5$ complex terminates abruptly just above the Fermi level because the upper $d$-band is nearly dispersionless there, whereas the hybridized TrDOS recovers after a shallow minimum and rises again.

\begin{figure*}[htbp]
     \centering
     \includegraphics[width=1.0\textwidth]{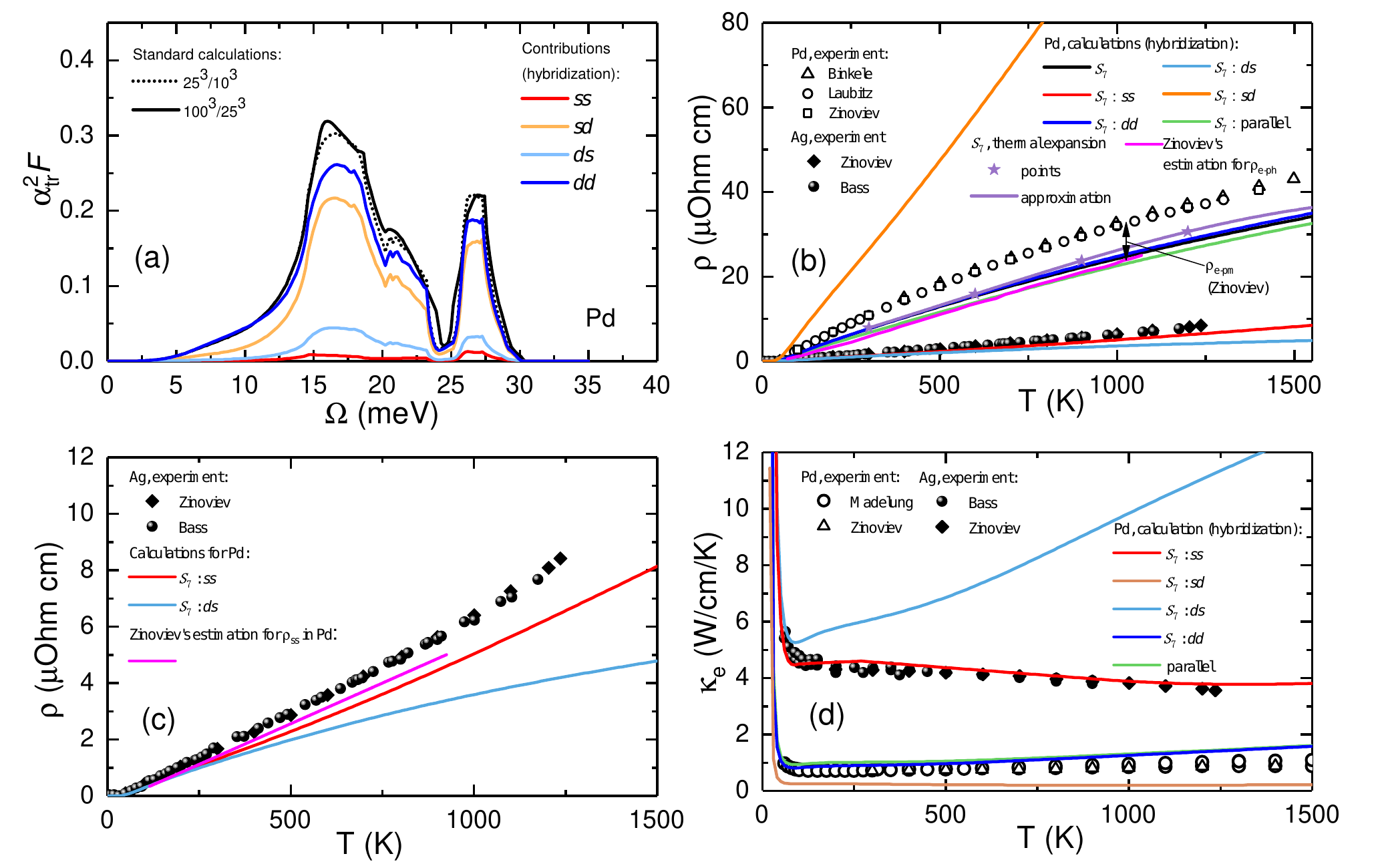}
     \caption{Electron-phonon transport properties of Pd. (a)~Allen's transport spectral function $\alpha_\mathrm{tr}^2F(1,1,\varepsilon_F,\Omega)$ in the \enquote{hybridized} approach: $s$-$s$ (red), $s$-$d$ (orange), $d$-$s$ (light blue), $d$-$d$ (blue); black lines show the unhybridized total on two different $\textbf{k}/\textbf{q}$ grids. (b)~Electrical resistivity versus temperature: experimental data for Pd (hollow triangles~\cite{binkele1986significance}, circles~\cite{laubitz1972high}) and Ag (diamonds~\cite{zinoviev1989}, circles~\cite{bass1982landolt}). Our calculations results are given in the same color map as for TSFs, thermal expansion is given in purple stars and purple line is the smooth approximation. (c)~Low-temperature zoom comparing Ag data and Zinoviev's~\cite{zinoviev1989} estimate for Pd. (d)~Electronic thermal conductivity; experimental data for Pd from Refs.~\cite{madelung1991thermal} (circles) and \cite{zinoviev1989} (triangles).}
     \label{fig:a2f_resist_thermal_hybrid_Pd}
\end{figure*}

The transport spectral function for Pd is shown in Fig.~\ref{fig:a2f_resist_thermal_hybrid_Pd}(a). 
All four orbital-resolved components reproduce profile obtained without accounting for hybridization: a primary peak near 16~meV and a secondary peak around 27~meV from the longitudinal acoustic branch. 
The $d$-$d$ term dominates, closely matching the total unhybridized spectrum---a direct consequence of the large $t_{2g}$-peak of $d$-states at the Fermi level. 
The second-largest contribution comes from $s$-$d$ transitions; it exceeds both the $s$-$s$ and $d$-$s$ channels and therefore produces the pronounced asymmetry expected in the Mott picture. 
Because the transport DOS of $s$-like states near the chemical potential is small while that of $d$ states is high, the $s$-$d$ channel acquires disproportionate weight in the transport coefficients and yields the large resistivity shown in orange in Fig.~\ref{fig:a2f_resist_thermal_hybrid_Pd}(b).

The $s$-$s$ channel provides the weakest weight in TSF, yet it remains visible in the temperature-dependent resistivity Fig.~\ref{fig:a2f_resist_thermal_hybrid_Pd}(b,c) because the $s$ TrDOS at the Fermi level is small. 
Its temperature dependence differs notably from the $d$-contributions: while $\rho_{dd}$ and $\rho_{ds}$ display the characteristic concave curvature at high $T$, $\rho_{ss}$ is almost linear and even slightly convex. 
Its calculated magnitude is close to the experimental data on resistivity of pure Ag, consistent with Zinoviev's theoretical estimate for Pd up to 1000~K~\cite{zinoviev1989}.

Although the bare $s$-$s$ term is modest, the full $s$-channel is dominated by $s$-$d$ scattering, whose magnitude exceeds the scale of Fig.~\ref{fig:a2f_resist_thermal_hybrid_Pd}(b). 
This strongly supports the Mott model: the $s$-$d$ resistivity alone is already about twice the experimental low-temperature slope of Pd and remains above the measured curve as the temperature increases. 
By contrast, the $d$-channel contributes far less; its leading $d$-$d$ component (blue curve) is essentially identical to the total resistivity calculated without hybridization, in line with the discussion of orbital-resolved TSFs above in Fig.~\ref{fig:a2f_resist_thermal_hybrid_Pd}(a).

Adding the two channels in parallel through Eqs.~\eqref{eq:rho_parallel}--\eqref{eq:rho_d_chan} gives the green curve in Fig.~\ref{fig:a2f_resist_thermal_hybrid_Pd}(b), which is very close to the standard Allen's method result. 
This confirms that the \enquote{hybridized} approach for resistivity decomposition is internally consistent: a proper parallel summation recovers the resistivity without electronic weighting factors. Figure~\ref{fig:a2f_resist_thermal_hybrid_Pd}(b) also contains experimental resistivity data for solid Pd up to melting, together with Zinoviev’s~\cite{zinoviev1989} theoretical estimates. Our standard Allen calculation performed without state weighting along the 300~K isochore yields the black curve, which sits roughly $10 \mu$Ohm cm below the measured values. Including thermal expansion via the Touloukian equation of state~\cite{touloukian1975thermalexpansion} partially closes this gap as the melting point is approached. The remaining offset is most naturally ascribed to paramagnon scattering, a contribution known to be strong in nearly magnetic metals such as the actinides and platinum and especially pronounced in Pd~\cite{Jullien1974, BEALMONOD19731707, MILLS1973679, kubler2021theory}. Because the Allen formalism treats only electron-phonon scattering, this contribution is absent in our calculations. Zinoviev’s estimate of the purely electron-phonon contribution (purple line, up to 1100~K) is also plotted; it is in an excellent agreement with our \textit{ab initio} results.

Figure~\ref{fig:a2f_resist_thermal_hybrid_Pd}(d) shows the electronic thermal conductivity. 
The hierarchy of contributions is reversed relative to the resistivity plot: the $s$-$d$ term is vanishingly small, whereas the $d$-$s$ contribution is the largest and develops an upturn above 500~K that is absent in all other channels. 
The $d$-$d$ thermal conductivity coincides with both the standard Allen's method result (not shown explicitly to make designations easier) and the parallel summation obtained from Eqs.~\eqref{eq:kappa_parallel} and \eqref{eq:kappa_both_chan}, and it agrees with the experimental data for Pd up to about 600~K; the subsequent discrepancy is attributed to thermal expansion, omitted here. 
The $s$-$s$ conductivity lies between the $d$-$s$ and $d$-$d$ values and again tracks the experimental data for Ag closely, confirming Zinoviev's estimation for the $s$-subgroup contribution in the thermal case as well.

\subsection{Mo}\label{subsec:Mo}

\begin{figure*}[htbp]
     \centering
     \includegraphics[width=1.0\textwidth]{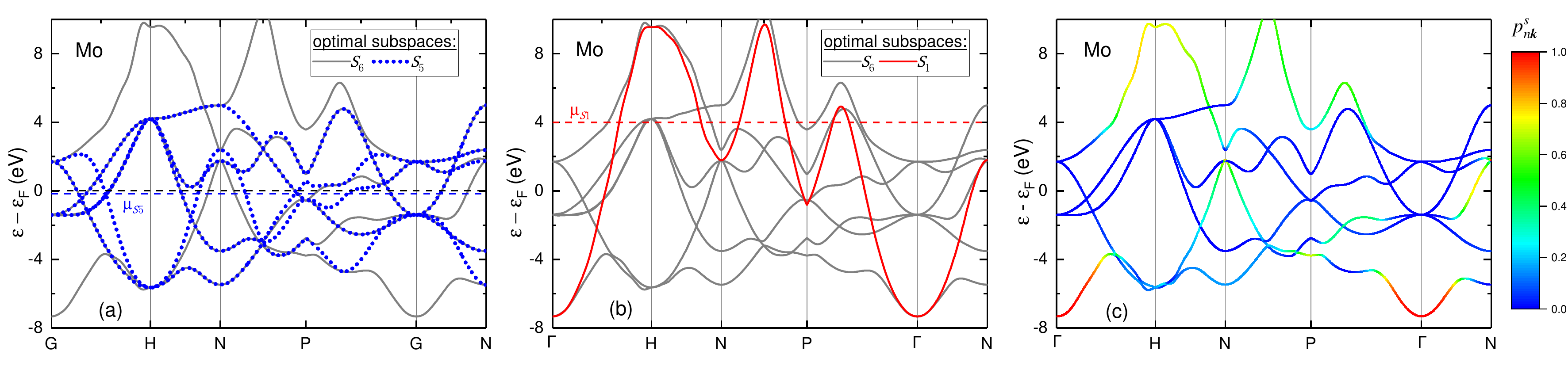}
     \caption{Comparison of \enquote{direct} and \enquote{hybridized} approaches for band structure of Mo. (a) Gray line denotes band structure obtained from $\mathcal{S}_6$ subspace, blue dotted line is for the $d$-complex obtained from $\mathcal{S}_5$. (b) The results for $\mathcal{S}_1$ subspace obtained as a complement of $\mathcal{S}_5$ are shown in red. (c) The same $\mathcal{S}_6$ bandstructure but colored according to orbital Wannier-projected weight taken over one $s$-type orbital.}
     \label{fig:direct_bands_Mo}
 \end{figure*}

The $d$-complex of the optimal $\mathcal{S}_5$ subspace is compared with the full $\mathcal{S}_6$ bands of bcc Mo in Fig.~\ref{fig:direct_bands_Mo}(a).
The obtained $d$-complex in Fig.~\ref{fig:direct_bands_Mo}(a) is quite close to the more complex hybridized band structure from $\mathcal{S}_6$ subspace of Mo in the vicinity of Fermi level. Note the position of Fermi level for separate $d$-complex shown in blue dashed horizontal line in $\mathcal{S}_5$ just a little lower the one obtained in $\mathcal{S}_6$ set as the energy zero (the number of $d$-electrons equals $N_d = 4.9$ in \enquote{hybridezed} approach for Mo). Most significant discrepancies of the two band structures presented in Fig.~\ref{fig:direct_bands_Mo} are seen along $\Gamma$-$H$ segment (one upper band of the doubly degenerate state in $\Gamma$ tends to decrease until it reaches $H$ instead of rising as in $\mathcal{S}_6$), around $N$-point at about $2$~eV above $\varepsilon_F$ and along $N$-$P$ and $P$-$G$ paths.

\begin{figure}[b]
     \centering
     \includegraphics[width=0.51\textwidth]{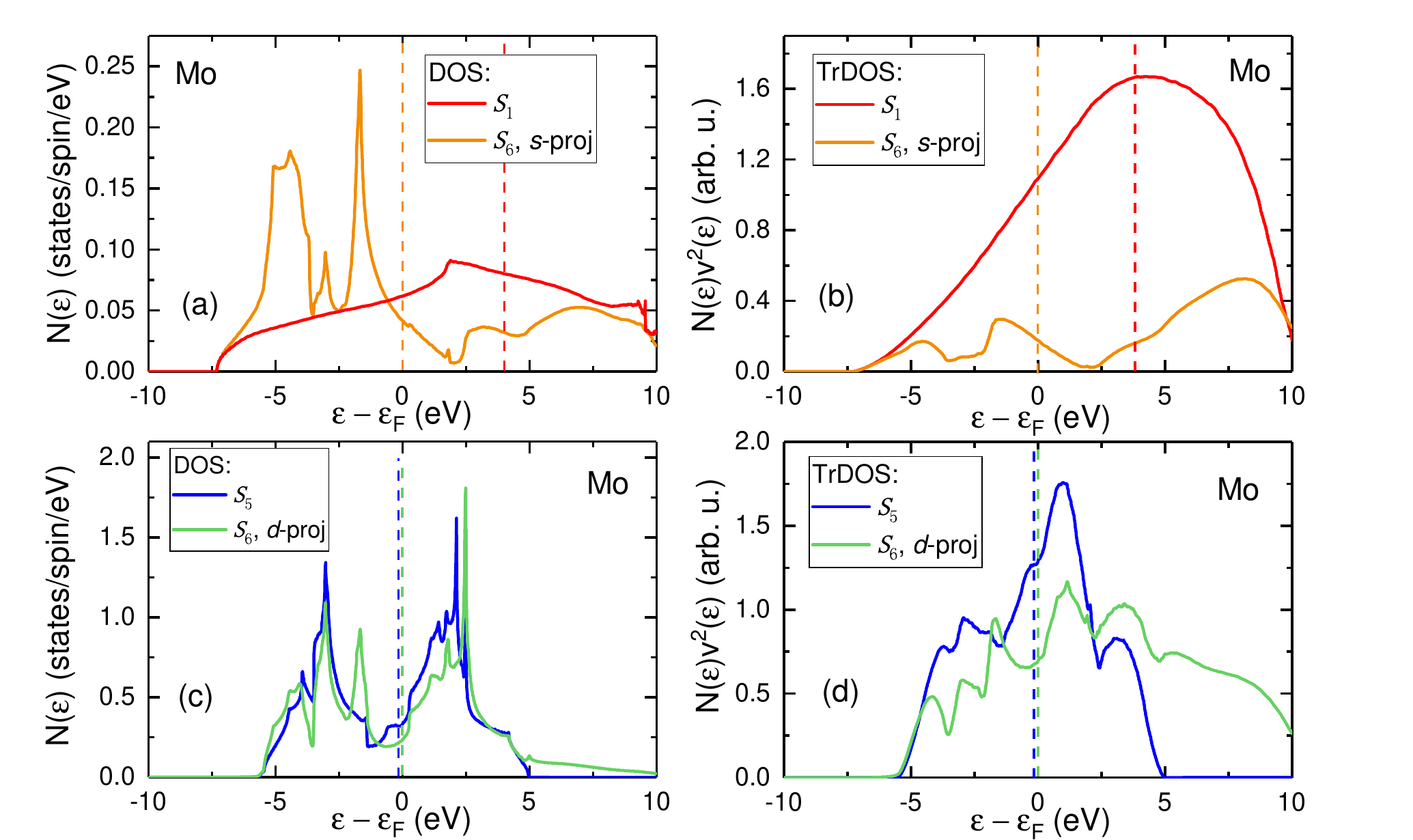}
     \caption{Comparison of DOS and TrDOS for Mo. All designations are the same as were used for Pd in Fig.~\ref{fig:DOS_Pd_direct_vs_hybrid}.}
     \label{fig:DOS_Mo_direct_vs_hybrid}
 \end{figure}

The wide $s$-band extracted from $\mathcal{S}_1$ as the complement of $\mathcal{S}_5$ is shown in red in Fig.~\ref{fig:direct_bands_Mo}(b). Its Fermi level is shifted at $4$~eV above the corresponding value in $\mathcal{S}_6$ to to contain the correct number of $s$-electrons which equals $N_s = 1.1$ as obtained from \enquote{hybridized} approach for Mo. As previously obtained for fcc Pd upon extraction of the $s$-like band, excellent agreement is observed here between the bottom of the $\mathcal{S}_1$ band and the lower band of $\mathcal{S}_6$ in the vicinity of the 
$\Gamma$-point. However, an important distinction from the fcc case is the presence of characteristic regions where the $\mathcal{S}_1$ band exhibits extrema: notably, at the $P$-point it lies in close proximity to the Fermi level of $\mathcal{S}_6$, and at the $N$-point another minimum is located approximately $2$~eV above the first one. These features correlate well with the band structure of the $\mathcal{S}_6$ subspace, color-coded according to the $s$-state weight along the same reciprocal-space path, as shown in Fig.~\ref{fig:direct_bands_Mo}(c). As in the case of fcc Pd, the most intense red hue is observed in regions where the 
$s$-band obtained within the \enquote{direct} approach intersects with the $\mathcal{S}_6$ bands, indicating the maximum contribution of 
$s$-type electrons in these regions. Conversely, as the $\mathcal{S}_6$ bands and the $\mathcal{S}_1$ band diverge, the weight of the $d$-states becomes increasingly dominant. Unlike palladium, which is characterized by a pronounced $d$-peak at the Fermi level, molybdenum has much fewer density of states at $\varepsilon_F$, see Fig.~\ref{fig:DOS_Mo_direct_vs_hybrid}(c). Nevertheless, $d$-state contribution still remains dominant. As a result, Fig.~\ref{fig:direct_bands_Mo}(c) clearly demonstrates a significant fraction of states with an $s$-to-$d$ weight ratio close to unity, indicating strong hybridization.

The DOS and TrDOS are compared in Fig.~\ref{fig:DOS_Mo_direct_vs_hybrid}.
For the $s$-states in panels~(a,b), the pattern mirrors that of Pd: the direct $\mathcal{S}_1$ result follows the free-electron square-root law, whereas the $\mathcal{S}_6$ projection onto one single MLWF yields a structured DOS with several peaks below $\varepsilon_F$ and a very small value right at the Fermi level.

\begin{figure*}[htbp]
     \centering
     \includegraphics[width=1.0\textwidth]{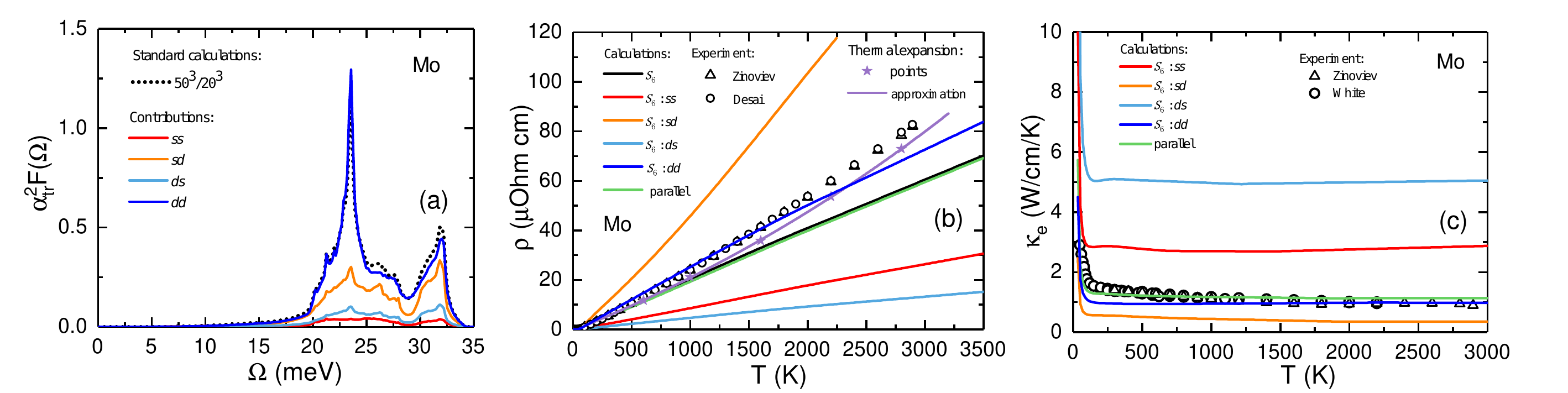}
     \caption{Transport properties for Mo. (a) Contributions to transport spectral function. All designations are the same as were used for Pd in Fig.~\ref{fig:a2f_resist_thermal_hybrid_Pd}. (b) Temperature dependence of electrical resistivity: all calculation data are shown is the same colors as for Pd in Fig.~\ref{fig:a2f_resist_thermal_hybrid_Pd}; experimental data are taken from~\cite{zinoviev1989} (triangles) and~\cite{desai1987thermodynamic} (circles). (c) Temperature dependence of electron thermal conductivity: same designations for calculations; experimental data are given in triangles~\cite{zinoviev1989} and circles~\cite{madelung1991thermal}.}
     \label{fig:a2f_resist_thermal_hybrid_Mo}
 \end{figure*}

Regarding the density of states of $d$-electrons, both the \enquote{direct} and \enquote{hybridized} approaches yield similar results, as in the case of Pd. However, for Mo, a characteristic minimum in the $d$-state density of states is observed in the vicinity of Fermi energy. Consequently, the contribution of $s$-$s$ transitions to the transport properties of Mo is significantly higher than that in Pd (see Fig.~\ref{fig:a2f_resist_thermal_hybrid_Mo}(b,c) below).

The magnitude of the transport density of states for $s$-electrons obtained through direct band extraction, as in the case of Pd (Fig.~\ref{fig:DOS_Mo_direct_vs_hybrid}(b)), significantly exceeds the corresponding value obtained within the \enquote{hybridized} approach, further emphasizing the fundamental difference in the kinetic properties of this group of electrons between the two approaches. This distinction is also noticeable for the 
$d$-states in Fig.~\ref{fig:DOS_Mo_direct_vs_hybrid}(d), although it is less pronounced.

Fig.~\ref{fig:a2f_resist_thermal_hybrid_Mo}(a) presents various orbital resolved TSFs at the Fermi level $\alpha^2_\mathrm{tr}F(1, 1, \varepsilon_F, \Omega)$ obtained within the \enquote{hybridized} approach. The main qualitative findings are the same as in the case of Pd, confirming the generality of the assumptions made in the discussion of the parallel addition rule for resistivities following eq.~\eqref{eq:rho_d_chan}. Thus, the contribution of $d$-$d$ transitions is found to be very close to the results of calculations performed without accounting for hybridization. The second-largest contribution in the $d$-channel, arising from $d$-$s$ scattering, is significantly smaller. In the 
$s$-channel, the dominant contribution again comes from transitions from initial $s$- to final $d$-states, which is consistent with the Mott model. The remaining $s$-$s$ contribution is the smallest when comparing the spectral functions; however, it remains quite noticeable in the resistivity plot shown in Fig.~\ref{fig:a2f_resist_thermal_hybrid_Mo}(b).

Our calculations of the temperature dependence of the electrical resistivity, performed within the standard Allen approach along the isochore corresponding to the equilibrium volume at 300 K, are shown by the black line in Fig.~\ref{fig:a2f_resist_thermal_hybrid_Mo}(b). Good agreement with experiment is observed up to temperatures of approximately $800$~K, above which the discrepancy increases significantly. Accounting for thermal expansion via equation of state~\cite{touloukian1975thermalexpansion} substantially improves the agreement with the experimental data. The contributions from various scattering events are presented for the case without thermal expansion. First of all, in this case, a nearly linear temperature dependence of all contributions (without accounting for thermal expansion) is observed. This is attributed to the absence of a characteristic peak in the density of states of molybdenum, in contrast to palladium. At the same time, the $\rho_{ss}$ contribution turns out to be significantly more pronounced than it is in Pd, due to the smaller difference in the densities of states of electrons of the two orbital types noted above. The magnitude of $\rho_{ds}$ contribution is approximately half that of $\rho_{ss}$. Conversely, $\rho_{sd}$ contribution turns out to be the largest, exceeding the experimental resistivity curve of molybdenum by nearly a factor of two. Finally, the result of the parallel addition of resistivities in the two scattering channels nearly perfectly reproduces the calculations performed within the standard Allen method.

Fig.~\ref{fig:a2f_resist_thermal_hybrid_Mo}(c) presents the calculated contributions to the electronic thermal conductivity of Mo without accounting for thermal expansion. Good agreement is observed between the calculations for $\rho_{dd}$, parallel summation of the two channels and the experimental data over a wide temperature range. The results obtained by the standard Allen's method are almost identical to the parallel addition result and are not shown. It is also worth noting that in Mo, all partial contributions exhibit a weak temperature dependence starting from approximately $300$~K, while the increase in thermal conductivity as the temperature approaches zero is accurately reproduced by our calculations. The smallest contribution, arising from the $s$-$d$ and $d$-$d$ scattering channels, is close to the result of the calculation without hybridization, whereas the largest contribution is associated with $d$-to-$s$ transitions. The remaining $s$-$s$ contribution occupies an intermediate position between the $d$-$s$ and $s$-$d$ channels.

\section{Conclusions}\label{sec.conclusion}

In conclusion, we have proposed and implemented the \enquote{hybridized} approach for calculating transport properties of crystalline metals, based on the maximally localized Wannier function formalism and the Allen's method for solving kinetic equation. We have demonstrated that the use of  disentanglement windows and specific arrangements of trial orbitals ensures excellent localization and an accurate reproduction of the band structure compared to direct DFT calculations. The method allows for the correct calculation of partial densities of states and orbital-resolved transport spectral functions, paving the way for a detailed analysis of the contributions of various electronic subgroups to kinetic phenomena in solids.

We demonstrated the applicability of this approach for two cubic transition metals, namely fcc Pd and bcc Mo, revealing fundamental flaws in the simplified model of direct band extraction as originally proposed by Souza, Marzari, and Vanderbilt (SMV) for fcc metals. The direct isolation of $s$- or, 
$d$-like bands leads to an incorrect description of the corresponding kinetic properties, whereas the \enquote{hybridized} approach properly accounts for inter-orbital mixing in reciprocal space. Furthermore, for the bcc lattice, we found that the $\mathcal{S}_6$ configuration is perfectly applicable for projecting onto $s$-states, describing electronic and EPC properties nearly as good as the $sp^3d^2$ setup.

Although Pd and Mo possess vastly different electronic properties (e.g., density of states and $s$/$d$ electron ratios at Fermi level), the physical conclusions derived within the \enquote{hybridized} approach prove to be universal. For both metals, the parallel addition of scattering channels yields results in nearly perfect agreement with full calculations based on the standard Allen's method for both electrical resistivity and electronic thermal conductivity. Furthermore, the $d$-$d$ contribution obtained via the \enquote{hybridized} approach closely matches the standard calculation without hybridization. This consistency strongly validates the correctness of decomposing the transport properties into specific orbital scattering channels. Within this universal framework, we also clarified specific material features: due to the absence of a characteristic $d$-peak at Fermi level in Mo, the role of $s$-scattering is significantly higher than in Pd.

The proposed decomposition validates earlier theoretical estimates~\cite{fradin1974,pinski1978,zinoviev1989} that the $s$-$s$ contribution to the resistivity (electronic thermal conductivity) of Pd is comparable to the total resistivity (electronic thermal conductivity) of pure silver. Our calculated total resistivity is furthermore in excellent agreement with Zinoviev’s~\cite{zinoviev1989} estimate for contribution associated with EPC in Pd. The remaining discrepancy with experiment is attributed to electron–paramagnon scattering, a mechanism necessarily strong in nearly magnetic metals such as palladium, and not considered in this paper.

We also note that accounting for thermal expansion allows for much better agreement with experimental data on resistivity up to high temperatures for both considered metals. Thus, the proposed method not only eliminates the artifacts inherent to direct band extraction but also provides a powerful and universal tool for the microscopic analysis of electron scattering mechanisms in transition metals with complex band structures.
\newline
\section{Acknowledgements}

This work is supported by Russian Science Foundation (grant No. 26-19-00980).

\bibliography{Electronic_Contributions_PRB_1}

\end{document}